\documentclass[aps,prl,onecolumn,showpacs,superscriptaddress,groupedaddress]{revtex4}
\usepackage{dcolumn}   
\usepackage{bm}        
\usepackage{amssymb}   

\usepackage{graphicx}
\usepackage{subfig}
\usepackage{color}
\usepackage{epstopdf}

\begin{document}

\author{I. J. Souza}
\email{itacy@fisica.ufmt.br}
\author{M. Godoy}
\email{mgodoy@fisica.ufmt.br}
\author{A. S. de Arruda}
\affiliation{Instituto de F\'isica, Universidade Federal de Mato Grosso, 78060-900, Cuiab\'a, Mato Grosso, Brazil.}
\email{aarruda@fisica.ufmt.br}
\author{T. M. Tunes}
\affiliation{Faculdade de Engenharia, Universidade Federal do Mato Grosso, 78060-900, V\'arzea Grande, Mato Grosso, Brazil.}
\email{thiagotunes@fisica.ufmt.br}

\title{Critical behavior of the spin-3/2 Blume Capel quantum model with two random transverse single-ion anisotropies}

\date{\today}

\begin{abstract}

Using  the approach based on the Bogoliubov inequality for free energy, we have studied the magnetic properties of the spin-3/2 Blume-Capel quantum model in the presence of the random transverse single-ion anisotropy (RTSIA). By  analyzis of the phase diagrams in $\tau \times \varphi_{x} $, $m \times \varphi_{x}$  and $m \times \tau$ plans, we have obtained information about how the presence of anisotropy affects the properties of the system. We have shown that the RTSIA  is responsible for changing the topology of the phase diagrams, compared to the pure model, leading to the appearance tricritical point (TCP).

\end{abstract}

\pacs{}
\maketitle

\section{Introduction}

The interest in knowing the magnetic properties of materials began in the last century and, until today,  it has attracted a large number of researchers who seek to understand the physics behind magnetic phenomena. In this context, Pierre Curie~\cite{curie1}, through experimental observations, was able to differentiate the magnetic properties of the ferromagnetic materials from those of paramagnetic materials, leading Lenz and Ising to propose the first many-body model to describe ferromagnetism~\cite {ising}, which became known as Ising model. This model initially failed to describe phase transitions in ferromagnetic materials. This failure was first cited by Heisenberg in his work given by Ref.~\cite{heisen1}. But later, with Pierls  arguments~\cite{pierls} and  with the exact solution  in two dimensions obtained by Onsager~\cite{ons1}, the Ising model was definitely recognized as one of the most important models of condensed matter and it is today the most studied models in magnetism~\cite{cem}.

The Ising model is basically a two-state model, namely spin-up and spin-down. Over time, to try to extract information about the physics that is responsible for phase transitions in magnetic materials, it was necessary to modify the model by incorporating more states, as well as other types of interactions, such as crystalline anisotropy, and disorders. One of the most important modifications was made by Blume~\cite{blu} and by Capel~\cite{cap}. Both, independently,  introduced  spins-1 and also included the single-ion anisotropy, which became known as the Blume-Capel (BC) model.  The BC model is  the simplest lattice model that can exhibit a tricritical point (TCP).  Another version of the BC model, with spin-3/2,  was introduced to explain the phase transitions in $DyVO_4 $~\cite{blu2, cook1, cook2, sayeta} and the critical properties of certain mixtures of ternary fluids~\cite{muka}. This model is very useful for describing disordered systems as long as random single-ion anisotropy~\cite{k1} is incorporated, random magnetic fields~\cite{cdd, a1, a2,  ku},  and disorders in the couplings between the spins~\cite{a3}. Other disordered systems  are formed by two sublattices with different spins that are subjected to random single-ion anisotropy~\cite{kenz, ita1, ze1, ze2} and random magnetic field~\cite{w1}.

Another modification made to the BC model is the introduction of a transverse single-ion anisotropy, which transforms the classical model into a quantum model. Its effects for the spin-1 models  have already been quite investigated~\cite{sou1,plak3, rir1}. On the other hand, the spin-3/2 version has not been much explored, only a few works have been published~\cite{duc, erhan}.The BC model has been explored by several methods, such as, two-spin cluster~\cite{lee}, variational methods~\cite{man}, constant coupling approximation~\cite{tana}, Monte Carlo simulations~\cite{lan1,bonfim,nb, boca}, finite-size scaling~\cite{beale, alca, feli, geh}, renormalization group methods \cite{berker,bonfim2,suza} and effective field approximation~\cite{kan1,jiang,chak,fiti}.

In this paper the interest is to study the phase diagram of the BC model with random transverse single-ion anisotropy, i.e., we would like to understand how transverse  anisotropy works in the phase diagram topology, and thermodynamic behavior of the spin-3/2 BC model, which were partially treated  in Ref.~\cite{erhan}.  So we will study the effects of the random transverse single-ion anisotropy (RTSIA) in the phase diagram and the thermodynamic properties of the spin-3/2  BC quantum model.  This study is a continuation of a previous work~\cite{clau} which performed the same study for the case spin-1  BC quantum model. In this work, we have used the approach the mean-field theory based on the Bogoliubov inequality for the Gibbs free energy,  also the Landau theory to obtain the second-order phase transition lines and of the TCPs.

The outline of this work is organized as follows: In Section II, we describe the model with two random transverse single-ion anisotropies (the  spin-3/2 BC quantum model),  and in addition, we present the calculations of Gibbs free energy, magnetization using  the mean-field  theory based on the Bogoliubov inequality for the Gibbs free energy. Also, the Landau theory to perform phase transition analyzes is presented, which consists of expanding the free energy in a power series in the order parameter.  In Section III, we describe our theoretical results and discussed several phase diagrams and magnetization curves. Finally, in Section IV we give a summary and our conclusions.
\section{The model and methodology}
\hspace{0.6cm}
To be able to investigate the effects of the RTSIA on the magnetic properties, in this paper, we have considered the spin-3/2 BC quantum model in the presence of two RTSIAs. This model is defined by Hamiltonian:
\begin{eqnarray}
{\mathcal H} =- J\sum\limits_{\left\langle {i,j} \right\rangle } {S_i^z}
{S_j^z} + \sum\limits_i^N {{D_i}{{\left( {S_i^z} \right)}^2}}  +
\sum\limits_i^N {{D_i^{\prime}}{{\left( {S_i^x} \right)}^2}}.
\label{1}
\end{eqnarray}

In this Hamiltonian, the first term represents the interaction between nearest-neighbour spins  located at $i$ and $j$ sites on the hyper-cubic lattice, where  the coupling interaction is taken to be positive, that is, $J> 0$, which favors ferromagnetic alignment. $N$ is the total number of sites of the lattice. The sum is performed over all the nearest-neighbour pairs $\left\langle {i,j} \right\rangle$. In the second (third) term is present the longitudinal (transversal) anisotropy, denoted by $D_i$ ($D_{i}^{\prime}$), where $S_i^{\alpha}$, with $\alpha = x$, $y$, $z$ are the components of the spin-3/2 operator at site $i$, and $S_i^z $  assume the eigenvalues $\pm 3/2$ and $ \pm 1/2$. In both terms, the sums are performed  over all the lattice sites.

The model above can be reduced to classical spin-3/2 Ising making  $D_{i}=D_{i}^{\prime}=0$. For the case, $D_{i}^{\prime}=0$, the spin-3/2 BC model is recovered. The critical properties of the spin-3/2 BC model in the absence of transverse anisotropy have been studied by Lima {\it et al}~\cite{lima}. Phase diagrams for the $S=1$ case have been obtained in two-dimensional and three-dimensional by using in a variety of techniques~\cite{landau1,bonfim,wortis,sa}.

In order to make easier the adjustment with experimental results, the Hamiltonian given by Eq. (\ref{1}) can be rewritten as follows,
\begin{eqnarray}
\mathcal{H} = - J\sum\limits_{\left\langle {i,j} \right\rangle } {S_i^z}
{S_j^z} - \sum\limits_i^N {{D^x_i}{{\left( {S^x_i} \right)}^2}}  -
\sum\limits_i^N {{D^y_i}{{\left( {S_i^y} \right)}^2}}, \label{2}
\end{eqnarray}
where we have used the spins identities $S(S+1)=(S^x)^2+(S^y)^2+(S^z)^2$ and $ S(S+1)=(S^x)^2+(S^y)^2+(S^z)^2$. In addition, we have defined the variables $D^x_i=(D - D^{\prime})$ and
$D^y_i = D$, where we have neglected irrelevant constants.

To obtain the phase diagrams of the spin-3/2 BC model, we have employed the approach based on the Bogoliubov inequality for free energy, which is given by
\begin{eqnarray}
{\mathcal F}\le{\mathcal F_0} + \left\langle {\mathcal H} - {\mathcal H_0}(\eta) \right\rangle_0 \equiv
\Phi \left( \eta  \right),
\label{3}
\end{eqnarray}
where $\mathcal H$ is the Hamiltonian we want to solve, $\mathcal H_0$ is a Hamiltonian of known solution that, {\bf as can be seen}, depends on the variational parameter $\eta$.
In this work, we have chosen the trial Hamiltonian below
\begin{eqnarray}
{\mathcal H_0}\left( \eta  \right) = - \eta \sum\limits_i {\left( {S_i^z}
\right)  } - \sum\limits_i {{D_i^x}{{\left( {S_i^x} \right)}^2}}  -
\sum\limits_i {{D_i^y}{{\left( {S_i^y} \right)}^2}}.
\label{8}
\end{eqnarray}

In Eq. (\ref{3}), $\mathcal F_0$ corresponds to the free energy of Hamiltonian $\mathcal H_0$, where $\langle...\rangle$ is a thermal average over the ensemble defined by ${\mathcal H_0}$. The variational principle of Bogoliubov  allows us to obtain a approximate free energy which is given for the minimum of $\Phi \left( \eta  \right)$  with respect to $\eta$, that is, ${\mathcal F}= \Phi \left( \eta  \right)_{min}$.

Substituting Eqs. (\ref{2}) and (\ref{8}) into Eq. (\ref{3}), and performing the convenient algebraic manipulations, we obtain  free energy per spin $\overline{\Phi}=\Phi/N$:
\begin{eqnarray}
\overline{\Phi}(\eta)&=&-\frac{1}{\beta}\ln{\left[2e^{\beta\left(\frac{\eta}{2}+\frac{5\delta}{4}\right)}\cosh{\left(\frac{\beta}{2}\sqrt{4\eta^2-4\eta\delta+\delta^2+3\Delta^2}\right)}\right.} \nonumber \\
&+& \left. 2e^{\beta\left(-\frac{\eta}{2}+\frac{5\delta}{4}\right)}\cosh{\left(\frac{\beta}{2}\sqrt{4\eta^2+4\eta\delta+\delta^2+3\Delta^2}\right)}\right] + \eta m-\frac{Jzm^2}{2},
\label{9}
\end{eqnarray}
where $z$ refers to the coordination number of lattice and  $m$ is the  magnetization per spin, given by
\begin{eqnarray}
m=\langle \frac{1}{N}\sum\limits_iS_i^z \rangle_o .
\label{10}
\end{eqnarray}

Also in Eq. (\ref{9}), in order to facilitate mathematical manipulation, we have done the following variable substitutions:
\begin{eqnarray}
\delta =({{{D^x}+{D^y}}}), \nonumber \\
\Delta =({{{D^x}-{D^y}}})\nonumber.
\label{11}
\end{eqnarray}

Minimizing Eq. (\ref{9}) with respect to $m$, that is, $\frac{\partial\overline{\Phi}_(\eta)}{\partial {m}}=0$, we obtain variational parameter, $\eta=Jzm$. In this way, we can write the Landau free energy, defined as $\Psi(m)=-\beta\overline{\Phi}(\eta)$, as follow
\begin{eqnarray}
\Psi(\eta)&=& \ln\left[2e^{\left(\frac{m}{2\tau}+\frac{5(\varphi_x+\varphi_y)}{4\tau}\right)}\cosh{\left(\frac{1}{\tau}\sqrt{m^2-m(\varphi_x+\varphi_y)+(\varphi_x^2-\varphi_x\varphi_y+\varphi_y^2)}\right)} \right. \nonumber\\
&+&  \left. 2e^{\left(-\frac{m}{2\tau}+\frac{5(\varphi_x+\varphi_y)}{4\tau}\right)}\cosh{\left(\frac{1}{\tau}\sqrt{m^2+m(\varphi_x+\varphi_y)+(\varphi_x^2-\varphi_x\varphi_y+\varphi_y^2)}\right)}\right] - \frac{m^2}{2\tau}, \label{13}
\end{eqnarray}
where $\beta = 1/k_{B}T$, and  the magnetization per site is given defined by
\begin{eqnarray}
m&=&\left[e^{\left(\frac{m}{2\tau}+\frac{5(\varphi_x+\varphi_y)}{4\tau}\right)}\cosh{\left(\frac{1}{\tau}\sqrt{m^2-m(\varphi_x+\varphi_y)+(\varphi_x^2-\varphi_x\varphi_y+\varphi_y^2)}\right)}\right.\nonumber\\
&+& \frac{2e^{(\frac{m}{2\tau}+\frac{5(\varphi_x+\varphi_y)}{4\tau})}\sinh{\left(\frac{1}{\tau}\sqrt{m^2-m(\varphi_x+\varphi_y)+(\varphi_x^2-\varphi_x\varphi_y+\varphi_y^2)}\right)}\left(\frac{2m}{\tau}-\frac{(\varphi_x+\varphi_y)}{\tau}\right)}{\frac{2}{\tau}\sqrt{m^2-m(\varphi_x+\varphi_y)+(\varphi_x^2-\varphi_x\varphi_y+\varphi_y^2)}} \nonumber\\
&-&e^{(-\frac{m}{2\tau}+\frac{5(\varphi_x+\varphi_y)}{4\tau})}\cosh{\left(\frac{1}{\tau}\sqrt{m^2+m(\varphi_x+\varphi_y)+(\varphi_x^2-\varphi_x\varphi_y+\varphi_y^2)}\right)}\nonumber\\
&+&\left.\frac{2e^{(-\frac{m}{2\tau}+\frac{5(\varphi_x+\varphi_y)}{4\tau})}\sinh{\left(\frac{1}{\tau}\sqrt{m^2+m(\varphi_x+\varphi_y)+(\varphi_x^2-\varphi_x\varphi_y+\varphi_y^2)}\right)}\left(\frac{2m}{\tau}+\frac{(\varphi_x+\varphi_y)}{\tau}\right)}{\frac{2}{\tau}\sqrt{m^2+m(\varphi_x+\varphi_y)+(\varphi_x^2-\varphi_x\varphi_y+\varphi_y^2)}}\right]\nonumber\\
&& \left[e^{(\frac{m}{2\tau}+\frac{5(\varphi_x+\varphi_y)}{4\tau})}\cosh{\left(\frac{1}{\tau}\sqrt{m^2-m(\varphi_x+\varphi_y)+(\varphi_x^2-\varphi_x\varphi_y+\varphi_y^2)}\right)}\right.\nonumber \\
&+&\left. 2e^{(-\frac{m}{2\tau}+\frac{5(\varphi_x+\varphi_y)}{4\tau})}\cosh{\left(\frac{1}{\tau}\sqrt{m^2+m(\varphi_x+\varphi_y)+(\varphi_x^2-\varphi_x\varphi_y+\varphi_y^2)}\right)}\right]^{-1}. \label{131}
\end{eqnarray}

In the above equations, we have defined the dimensionless parameters,
$$\tau = \frac{1}{{\beta Jz}},\,\,\,{\varphi_x}\equiv\frac{{{D_x}}}{{Jz}},\,\,\, {\varphi_y}\equiv\frac{{{D_y}}}{{Jz}}.$$

The influence that anisotropy plays in the system is investigated by the average over the disorder, defined as
\begin{eqnarray}
{\langle \Psi \rangle}_c=\int\int P(\varphi_x)P(\varphi_y)\Psi d\varphi_x d\varphi_y ,
\label{14}
\end{eqnarray}
and
\begin{eqnarray}
{\langle m \rangle}_c=\int\int P(\varphi_x)P(\varphi_y) m d\varphi_x d\varphi_y .
\label{15}
\end{eqnarray}
In Eqs. (\ref{14}) and (\ref{15}), $P(\varphi_x)$ and $P(\varphi_y)$  are bimodal distributions that control the amount of anisotropy to which the system is subjected. $P(\varphi_x)$ and $P(\varphi_y)$ are given by,
\begin{eqnarray}
 P(\varphi_x)=p\delta(\varphi_x-\delta_x)+(1-p)\delta(\varphi_x),\nonumber \\
 P(\varphi_y)=q\delta(\varphi_y-\delta_y)+(1-q)\delta(\varphi_y),
 \label{16}
\end{eqnarray}
where $q$ and $p$ assume values ranging from $0$ to $1.0$.  The first term, in both equations, represents the amount of spins that are under the influence of anisotropies action with  strength $\varphi_x$ and $\varphi_y$. In contrast, the second term expresses the amount of spins that are not influenced by anisotropies. Therefore, for the case $q =p =1.0$, all the spins of the system are under the influence of anisotropies. The pure case is defined when $p$ and $q$ assume the value equal to $0$, that is, when the system is free of anisotropies.
Calculating the Eqs. (\ref{14}) and (\ref{15}), we have obtained
\begin{eqnarray}
{\langle \Psi \rangle}_c &=& pq \ln\left[2e^{\left(\frac{m}{2\tau}+\frac{5(\varphi_x+\varphi_y)}{4\tau}\right)}\cosh{\left(\frac{1}{\tau}\sqrt{m^2-m(\varphi_x+\varphi_y)+(\varphi_x^2-\varphi_x\varphi_y+\varphi_y^2)}\right)}\right. \nonumber \\
&+& \left. 2e^{\left(-\frac{m}{2\tau}+\frac{5(\varphi_x+\varphi_y)}{4\tau}\right)}\cosh{\left(\frac{1}{\tau}\sqrt{m^2+m(\varphi_x+\varphi_y)+(\varphi_x^2-\varphi_x\varphi_y+\varphi_y^2)}\right)}\right] \nonumber \\
&+& p(1-q) \ln{\left[2e^{\left(\frac{m}{2\tau}+\frac{5(\varphi_x)}{4\tau}\right)}\cosh{\left(\frac{1}{\tau}\sqrt{m^2-m\varphi_x+\varphi_x^2}\right)}\right. }\nonumber \\
&+& \left. 2e^{\left(-\frac{m}{2\tau}+\frac{5(\varphi_x)}{4\tau}\right)}\cosh{\left(\frac{1}{\tau}\sqrt{m^2+m\varphi_x+\varphi_x^2}\right)}\right]\nonumber \\
&+& q(1-p) \ln{\left[2e^{\left(\frac{m}{2\tau}+\frac{5(\varphi_y)}{4\tau}\right)}\cosh{\left(\frac{1}{\tau}\sqrt{m^2-m(\varphi_y)+(\varphi_y^2)}\right)}\right.} \nonumber \\
&+& \left. 2e^{\left(-\frac{m}{2\tau}+\frac{5(\varphi_y)}{4\tau}\right)}\cosh{\left(\frac{1}{\tau}\sqrt{m^2+m\varphi_y+\varphi_y^2}\right)}\right] \nonumber \\
&+& (1-p)(1-q)\ln{\left[4\cosh{\left(\frac{m}{\tau}\right)}\cosh{\left(\frac{m}{2\tau}\right)}\right]} - \frac{m^2}{2\tau} ,
 \label{17}
\end{eqnarray}
making the same way with magnetization
\begin{eqnarray}
{\langle m \rangle}_c=\int\int P(\varphi_x)P(\varphi_y) m d\varphi_x d\varphi_y , \label{38}
\end{eqnarray}
and
\begin{eqnarray}
\langle m  \rangle_c = pq A_1 + p(1-q) A_2 + q(1-p) A_3 + (1-p)(1-q) A_4,
\end{eqnarray}
where
\begin{eqnarray}
A_1 &=& \left[e^{\left(\frac{m}{2\tau}+\frac{5(\varphi_x+\varphi_y)}{4\tau}\right)}\cosh{\left(\frac{1}{\tau}\sqrt{m^2-m(\varphi_x+\varphi_y)+(\varphi_x^2-\varphi_x\varphi_y+\varphi_y^2)}\right)}\right.\nonumber\\
&+& \frac{2e^{\left(\frac{m}{2\tau}+\frac{5(\varphi_x+\varphi_y)}{4\tau}\right)}\sinh{\left(\frac{1}{\tau}\sqrt{m^2-m(\varphi_x+\varphi_y)+(\varphi_x^2-\varphi_x\varphi_y+\varphi_y^2)}\right)}\left(\frac{2m}{\tau}-\frac{(\varphi_x+\varphi_y)}{\tau}\right)}{\frac{2}{\tau}\sqrt{m^2-m(\varphi_x+\varphi_y)+(\varphi_x^2-\varphi_x\varphi_y+\varphi_y^2)}} \nonumber\\
&-& e^{\left(-\frac{m}{2\tau}+\frac{5(\varphi_x+\varphi_y)}{4\tau}\right)}\cosh{\left(\frac{1}{\tau}\sqrt{m^2+m(\varphi_x+\varphi_y)+(\varphi_x^2-\varphi_x\varphi_y+\varphi_y^2)}\right)}\nonumber\\
&+& \left.\frac{2e^{\left(-\frac{m}{2\tau}+\frac{5(\varphi_x+\varphi_y)}{4\tau}\right)}\sinh{\left(\frac{1}{\tau}\sqrt{m^2+m(\varphi_x+\varphi_y)+(\varphi_x^2-\varphi_x\varphi_y+\varphi_y^2)}\right)}\left(\frac{2m}{\tau}+\frac{(\varphi_x+\varphi_y)}{\tau}\right)}{\frac{2}{\tau}\sqrt{m^2+m(\varphi_x+\varphi_y)+(\varphi_x^2-\varphi_x\varphi_y+\varphi_y^2)}}\right]\nonumber\\
&\times & \left[e^{\left(\frac{m}{2\tau}+\frac{5(\varphi_x+\varphi_y)}{4\tau}\right)}\cosh{\left(\frac{1}{\tau}\sqrt{m^2-m(\varphi_x+\varphi_y)+(\varphi_x^2-\varphi_x\varphi_y+\varphi_y^2)}\right)}\right.\nonumber \\
&+&\left. 2e^{\left(-\frac{m}{2\tau}+\frac{5(\varphi_x+\varphi_y)}{4\tau}\right)}\cosh{\left(\frac{1}{\tau}\sqrt{m^2+m(\varphi_x+\varphi_y)+(\varphi_x^2-\varphi_x\varphi_y+\varphi_y^2)}\right)}\right]^{-1},
\end{eqnarray}
\begin{eqnarray}
A_2 &=& \left[e^{\left(\frac{m}{2\tau}+\frac{5\varphi_x}{4\tau}\right)}\cosh{\left(\frac{1}{\tau}\sqrt{m^2-m\varphi_x+\varphi_x^2}\right)}\right.\nonumber\\
&+& \frac{2e^{\left(\frac{m}{2\tau}+\frac{5\varphi_x}{4\tau}\right)}\sinh{\left(\frac{1}{\tau}\sqrt{m^2-m\varphi_x+\varphi_x^2}\right)}\left(\frac{2m}{\tau}-\frac{\varphi_x}{\tau}\right)}{\frac{2}{\tau}\sqrt{m^2-m\varphi_x+\varphi_x^2}} \nonumber\\
&-&e^{\left(-\frac{m}{2\tau}+\frac{5\varphi_x}{4\tau}\right)}\cosh{\left(\frac{1}{\tau}\sqrt{m^2+m\varphi_x+\varphi_x^2}\right)}\nonumber\\
&+&\left.\frac{2e^{\left(-\frac{m}{2\tau}+\frac{5\varphi_x}{4\tau}\right)}\sinh{\left(\frac{1}{\tau}\sqrt{m^2+m(\varphi_x+\varphi_y)+\varphi_x^2}\right)}\left(\frac{2m}{\tau}+\frac{\varphi_x}{\tau}\right)}{\frac{2}{\tau}\sqrt{m^2+m\varphi_x+\varphi_x^2}}\right]\nonumber\\
&\times & \left[e^{\left(\frac{m}{2\tau}+\frac{5\varphi_x}{4\tau}\right)}\cosh{\left(\frac{1}{\tau}\sqrt{m^2-m(\varphi_x+\varphi_y)+\varphi_x^2}\right)}\right.\nonumber \\
&+&\left. 2e^{\left(-\frac{m}{2\tau}+\frac{5\varphi_x}{4\tau}\right)}\cosh{\left(\frac{1}{\tau}\sqrt{m^2+m\varphi_x+\varphi_x^2}\right)}\right]^{-1},
\end{eqnarray}
\begin{eqnarray}
A_3 &=& \left[e^{\left(\frac{m}{2\tau}+\frac{5\varphi_y}{4\tau}\right)}\cosh{\left(\frac{1}{\tau}\sqrt{m^2-m\varphi_y+\varphi_y^2}\right)}\right.\nonumber\\
&+& \frac{2e^{\left(\frac{m}{2\tau}+\frac{5\varphi_y}{4\tau}\right)}\sinh{\left(\frac{1}{\tau}\sqrt{m^2-m\varphi_y+\varphi_y^2}\right)}\left(\frac{2m}{\tau}-\frac{\varphi_y}{\tau}\right)}{\frac{2}{\tau}\sqrt{m^2-m\varphi_y+\varphi_y^2}} \nonumber\\
&-&e^{\left(-\frac{m}{2\tau}+\frac{5\varphi_y}{4\tau}\right)}\cosh{\left(\frac{1}{\tau}\sqrt{m^2+m\varphi_y+\varphi_y^2}\right)}\nonumber\\
&+&\left.\frac{2e^{\left(-\frac{m}{2\tau}+\frac{5\varphi_y}{4\tau}\right)}\sinh{\left(\frac{1}{\tau}\sqrt{m^2+m\varphi_y+\varphi_y^2}\right)}\left(\frac{2m}{\tau}+\frac{\varphi_y}{\tau}\right)}{\frac{2}{\tau}\sqrt{m^2+m\varphi_y+\varphi_y^2}}\right]\nonumber\\
&\times & \left[e^{\left(\frac{m}{2\tau}+\frac{5\varphi_y}{4\tau}\right)}\cosh{\left(\frac{1}{\tau}\sqrt{m^2-m\varphi_y+\varphi_y^2}\right)}\right.\nonumber \\
&+&\left. 2e^{\left(-\frac{m}{2\tau}+\frac{5\varphi_y}{4\tau}\right)}\cosh{\left(\frac{1}{\tau}\sqrt{m^2+m\varphi_y+\varphi_y^2}\right)}\right]^{-1},
\end{eqnarray}
\begin{eqnarray}
A_4 &=&  \left[{\cosh\left(\frac{m}{\tau}\right)\sinh\left(\frac{m}{2\tau}\right)+\cosh\left(\frac{m}{2\tau}\right)\sinh\left(\frac{m}{\tau}\right)}\right]\nonumber\\
&\times & \left[\cosh\left(\frac{m}{\tau}\right)\cosh\left(\frac{m}{2\tau}\right)\right]^{-1}.
\label{18}
\end{eqnarray}

In general, the model that  we have studied, given by the Hamiltonian in Eq. (\ref{1}), exhibits a tricritical behavior, with first- and second-order phase transitions, for certain anisotropy values. In order to study these transitions in detail, we have expanded the Eq. (\ref{17}) as Landau-like expansion of the free energy, as
\begin{eqnarray}
\Psi(m)\simeq a_2m^2+a_4m^4+a_6m^6+ \cdots ,
\label{19}
\end{eqnarray}
where $a_2$, $a_4$, $a_6$ are coefficients depending on $\tau$, $\delta_{x}$, $\delta_{y}$, $p$, $q$, calculated by $ a_p=\frac{1}{p!}{\left(\frac{\partial^p\Psi}{\partial m_p}\right)}_{m=0}$.

We have computed the coefficients given above. However, because their expressions are extensive, we present in the appendix A all results in detail.

\section{Results}

In this section, we present our results and discuss the effects that RTSIA plays on the phase diagrams and thermodynamic properties of the spin-3/2 BC model. These results were obtained by solving numerically the Eqs. (12),   (14) and the coefficients of the Eq. (19).

In phase diagrams,  the second-order transition lines (continuous phase transition), in addition to the TCP, are obtained by analyzes of the coefficients found from the free energy expansion in Eq. (19). The continuous phase transition is characterized by breakdown symmetry of the order parameter $m$ (magnetization) close to the critical point, that is,  $m \rightarrow 0$. By numerical solution for $a_{2}=0$ and $a_{4} > 0$ provides the  second-order transition lines. The TCP defines the point at which the phase transition changes from second-order (continuous) to first-order  (discontinuous), and it is obtained by solving numerically $a_{2} =0$, $a_{4}=0$ and $a_{6}> 0$. The first-order lines are obtained from the comparison of the free energy for several solutions of the order parameter, given by Eq. (14).

\subsection{Phase diagrams in $\tau - \varphi_x$ plane}

We have started our analyzes by examining phase diagram in $\tau - \varphi_x$ plane  for $\varphi_{y}= 0.50$ as shown in Fig.~\ref{1}. We have varied $p$ from 0 to 1.0 and analyzed the cases $q=0$, $q = 0.50$, $q = 0.60$ and $q = 0.80$, which are presented in Figs.~\ref{1}(a), (b), (c) and (d), respectively.

For the case $q=0$,  in which our sample consists of the following configuration: the $N$ spins of the system are free from the action of the $\varphi_{y}$  anisotropy, that is, $\varphi_{y} = 0$. The anisotropy $ \varphi_x\not=0 $ will act on the $N$ spins depending on the value of the parameter $ p $, so $ p = 0 $ indicates that all spins are free from the influence  this anisotropy. The parameter $ p $ varies from 0 to 1.0, and when $p$ grows the number of spins in the system that are affected by this anisotropy also grows, as can be seen in Fig.~\ref{1}.  The solid (dotted) lines correspond to the lines second-order (first-order), and TCPs are represented by black dots. Then for  $p=0$ (pure case) we have obtained, as expected, that the critical temperature  does not change when $\varphi_x$ varies, and it is set at 1.25 (see Fig.~\ref{1}(a)). In contrast, as the spins begin to experience the effects of anisotropy ($p\not=0$), typical patterns of second- and first-order transition lines begin to emerge. For $0 \leq p < 0.95$, we have only observed second-order transition lines separating the ferromagnetic ($F$) from the paramagnetic ($P$) phases. This topology of the phase diagram is changed when we reach values of $p=0.95$ (95\% of spins are under the influence of anisotropy), where we begin to observe the existence of TCP separating the region with second-order phase transitions of the region with first-order phase transitions. For $p=0.95$ and $1.0$, the $\varphi_x$($\tau$) values for which TCP appears are of 0.80 (0.55) and 0.70 (0.67), respectively.
 As can be seen in Fig.~\ref{1}(a),  the system is sensitive to the presence of anisotropy $\varphi_{x}$, which plays a crucial role in second- and  first-order phase transitions, besides lowering  the critical temperature. Only when $95\%$ of the spins are under the influence of this anisotropy $\varphi_{x}$ the first-order transition appears. This case was treated within another context by E. Albayrak~\cite{erhan} who showed the phase diagram in two dimensions.

\begin{figure}[h]
\centering
\includegraphics[scale=0.60]{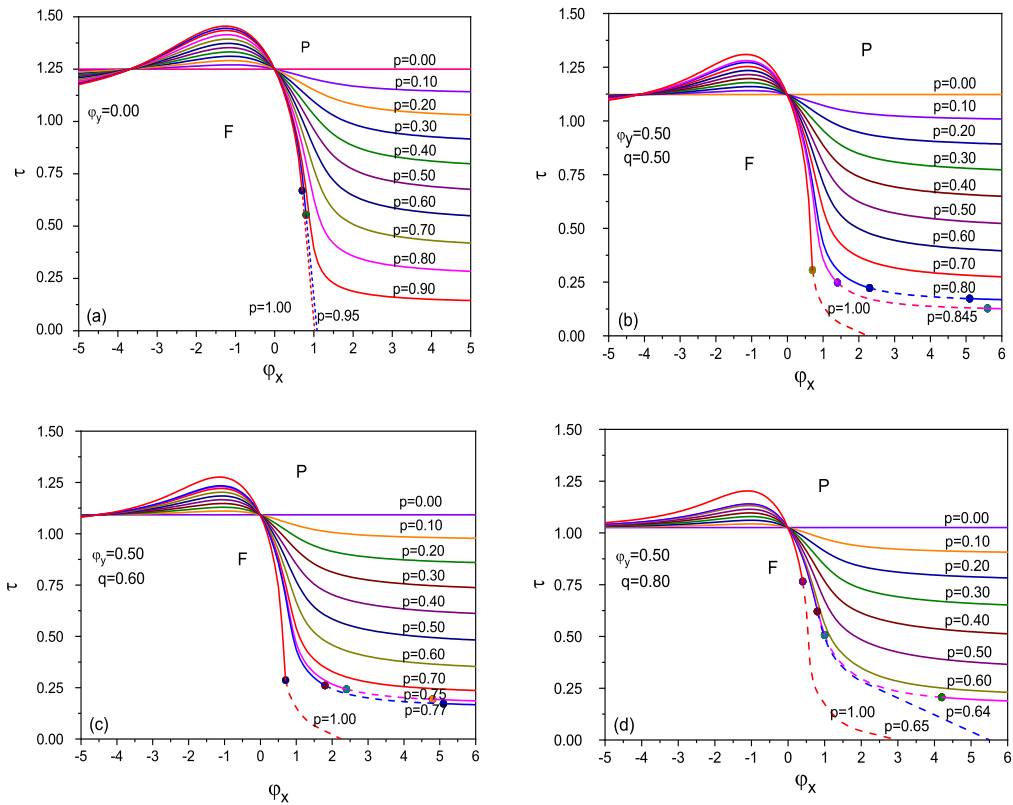}
\caption{(Color online) Phase diagram in $\tau - \varphi_x$ plane of the spin-3/2 BC quantum model with two RTSIA's for $\varphi_{y}=0.5$, different values of $p$ and $q$: (a) $q =0$, (b) $q=0.50$, (c) $q=0.60$ and (d) $q = 1.0$. $F$ and $P$ are the ferromagnetic and paramagnetic phases, respectively. The dots represent the TCP.}
\label{1}
\end{figure}

When we increase the number of spins subject to anisotropy $\varphi_{y}$, that is, the value of $q$ becomes greater (see in Fig.~\ref{1}(b) with  $q =0.5 $ (50\%))  this anisotropy makes the critical temperature to decrease (see case $p = 0$). Therefore, it induces the first-order transition for smaller values of $p$, i.e., $ p \geq 0.80$ for a range of $\varphi_x$ ($\tau$) between $2.30$ (0.22) and $5.10$ (0.17). Here, we can see that the transition again becomes second-order for large values of $\varphi_x$.  In this case,  the TPC are located only in regions of lower temperatures.

In Fig.~\ref{1}(c), we illustrated the results for $q = 0.60$. It is possible to observe that as we increase the number of spins that are under the influence of anisotropy $\varphi_{y}$, it has the same effect at $p = 0$ and the  first-order transition is observed for values smaller of $p$ which, in this case, occurs for $p\geq 0.75$. In the last case, as for $p = 0.77$, the first-order transition is only observed for a range of $\varphi_x$ ($\tau$) which corresponds to $2.40 - 4.80$ ($0.24-0.19$) and $1.80 - 5.10$ ($0.26-0.17$), respectively. Also here we see that the transition again becomes second-order for large values of $\varphi_x$.

Now, in Fig.~\ref{1}(d), we have considered the case where 80\% of spins are under the influence of anisotropy, which corresponds to $ q = 0.80 $. As can be seen in Fig.~\ref{1}(d),
the effect at $p = 0$ is quite pronounced, that is, there is a greater decrease in the critical temperature. In this case for values of $p < 0.64$ we have only observed second-order transition lines. When $p$ assumes the value of 0.64, the system displays a TCP. For this case, the first-order transition lines are observed for $\varphi_x$ ($\tau$) of $1.00-4.20$ ($0.506-0.206$). Here, we have seen that the transition again becomes second-order for large values of $\varphi_x$, however, with the increase of $p$ the first-order transition ends at $\tau=0$. Comparing with other cases, we see that the TCPs appear for smaller values of $p$ when we increase the value of $q$. This shows how the topology of the phase diagrams is mainly affected by the anisotropies $\varphi_{y}$, which introduced the tricritical behavior for smaller values of $p$.

\begin{figure}[h]
\centering
\includegraphics[scale=0.55]{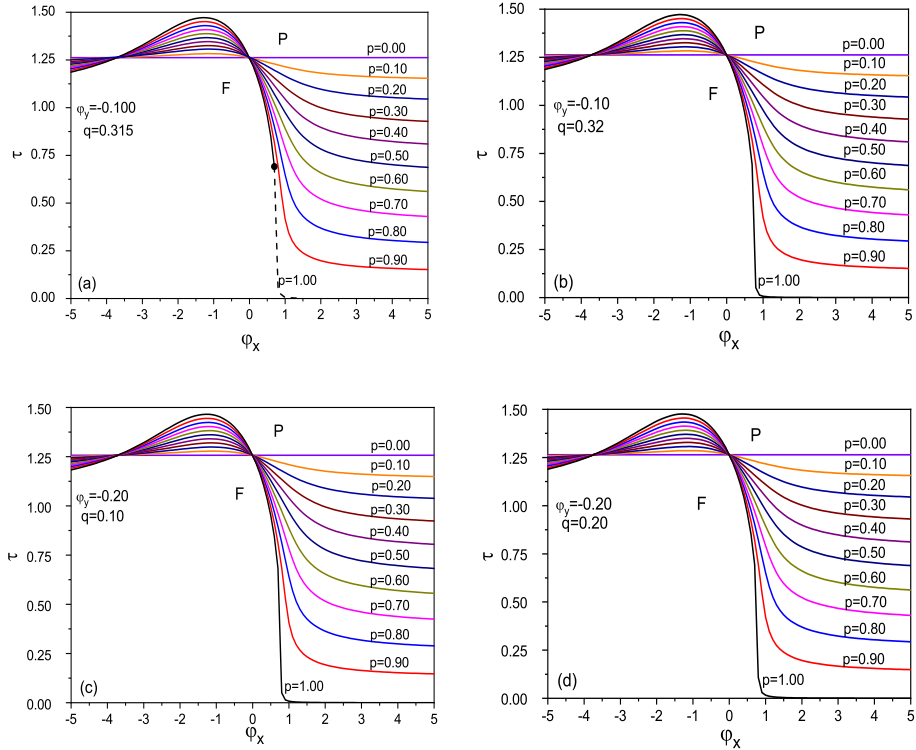}
\caption{(Color online) Phase diagram in $\tau - \varphi_x$ plane of the spin-3/2 BC model with two RTSIA's for $\varphi_{y}=-0.10$, and for different values of $p$ and $q$: (a) $q=0$, (b) $q=0.315$, (c) $q=0.32$ and $\varphi_{y}=-0.10$ with (d) $q=0$. $F$ and $P$ are the ferromagnetic and paramagnetic phases, respectively. Black points represents the TCP.}
\label{2}
\end{figure}

Now, in Fig.~\ref{2} we have considered the case where $\varphi_y$ assumes the negative values, that is,   $-0.10$ and $-0.20$.  We would like to emphasize that when $q = 0$ regardless of the value of $\varphi_{y}$ we will always be in the case of Fig.~\ref{1}(a). Thus, we start with a concentration of spins $q = 0.315$ $(31.5\%)$ under the action of anisotropy $\varphi_{y} = - 0.10$.  We observed that  the first-order transition lines only emerge for $p =1.0$, where all the spins are under influence of anisotropy given by $\varphi_x$. In order to investigate whether, in fact, $q =0.315 $ corresponds to the critical value from which the first-order lines are suppressed, we investigated the case where $q =0.32$ (see Fig.~\ref{3}(b)). We can observe that there is no presence of the first-order transition for any values of $p$ investigated. Therefore, for $\varphi_{y}=-0.10$, the presence of tricritical points is only observed for $q \leq 0.315$. As we have seen in Fig.~\ref{2}(b), for $\varphi_{y}=-0.10$, the first-order transition lines are present only for $q \leq 0.315$. Unlike this case, for $\varphi_{y}=-0.20$ (see Fig.~\ref{2}(c) and Fig.~\ref{2}(d)), we have not observed first-order transition lines for any value of $q$. Therefore, the reduction of the magnitude of $\varphi_{y}$ (in particular to negative values) is responsible for the appearance of first-order transition lines.  Therefore, we can conclude that $\varphi_{y}$ plays a fundamental role in the appearance of the first-order phase transition and when negative it increases the critical temperature.

\subsection{Magnetization curves}

In the previous section, we have seen how the introduction of anisotropy makes to modify the structure of the phase diagram of the pure BC model, leading to the appearance of tricritical behavior. In this section, to corroborate our results of the phase diagrams, we have obtained magnetization curves as a function of temperature $\tau$ and anisotropy $\varphi_x$ for the same values of $p$ and $q$. The magnetization curves were obtained by numerical solution of the Eq. (14).

In Fig.~\ref{1}(a), for the case $q=0$ and $\varphi_y=0$, we have seen that the TCP is not observed for $p<0.95$ for any value of $\varphi_x$, that is, the system does not display first-order transition lines. To confirm this picture, in Fig.~\ref{3}(a) we exhibited our results for magnetization versus temperature for $p =0.90$, $p=0.95$, and $p=1.00$. We have chosen some $\varphi_x$ values between $0.15$ and $1.00$ to show the appearance of first-order transition lines. By analyzes of the shape of curves (3), (4) and (5), we can observe a behavior characteristic of a second-order phase transition, where magnetization goes to zero continuously. We also can notice that even with an increase of $\varphi_x$,  the magnetization does not change its form. In contrast, the curves (1) $p =1.0$ and $\varphi_x=0.95$ and (2) $p =0.95$ and $\varphi_x=1.00$ shows a behavior characteristic of a first-order phase transition where there is a rapid variation of $m$  a function of temperature. These results confirm our results obtained for the phase diagrams in Fig.~\ref{1}(a). In Fig.~\ref{3}(b) we have shown the magnetization curve now as a function of $\varphi_x$. In this analyzes, we have fixed the temperature $\tau$. The curve (2) with $p=1.00$ and ($\tau =0.25$) shows a discontinuity in magnetization, that is, a first-order transition. When the temperature increases (see curve (1) with $p=1.00$ and ($\tau =1.00$)) the magnetization goes to zero continuously for a critical value of $\varphi_x$. These results are in agreement with that shown in Fig.~\ref{1}(a).

\begin{figure}[h]
\centering
\includegraphics[scale=0.38]{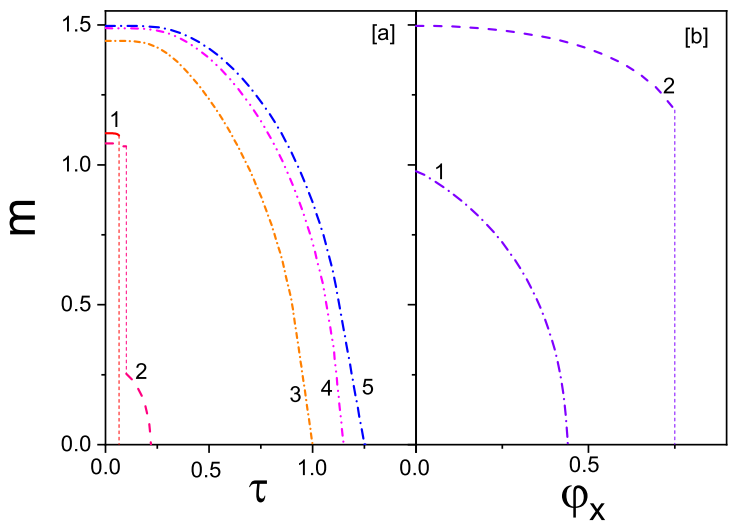}
\caption{(Color online) Behavior of magnetization $m$ versus temperature $\tau$ (a) and parameter $\varphi_x$ (b) of the spin-3/2 BC model with two RTSIAs for $q=0$ or $\varphi_{y}=0$.  Here, we have used the following set of parameters:  $(a) - (1)$ $p =1.0$ and $\varphi_x=0.95$, (2) $p =0.95$ and $\varphi_x=1.00$, (3) $p =0.95$ and $\varphi_{x}=0.50$, (4) $p =1.0$ and $\varphi_{x}=0.25$, (5) $p=0.90$ and $\varphi_{x}=0.15$.  $(b) - (1)$ $p=1.00$ and $\tau =1.00$, (2) $p=1.00$ and $\tau=0.25$.}
\label{3}
\end{figure}

\begin{figure}[h]
\centering
\includegraphics[scale=0.38]{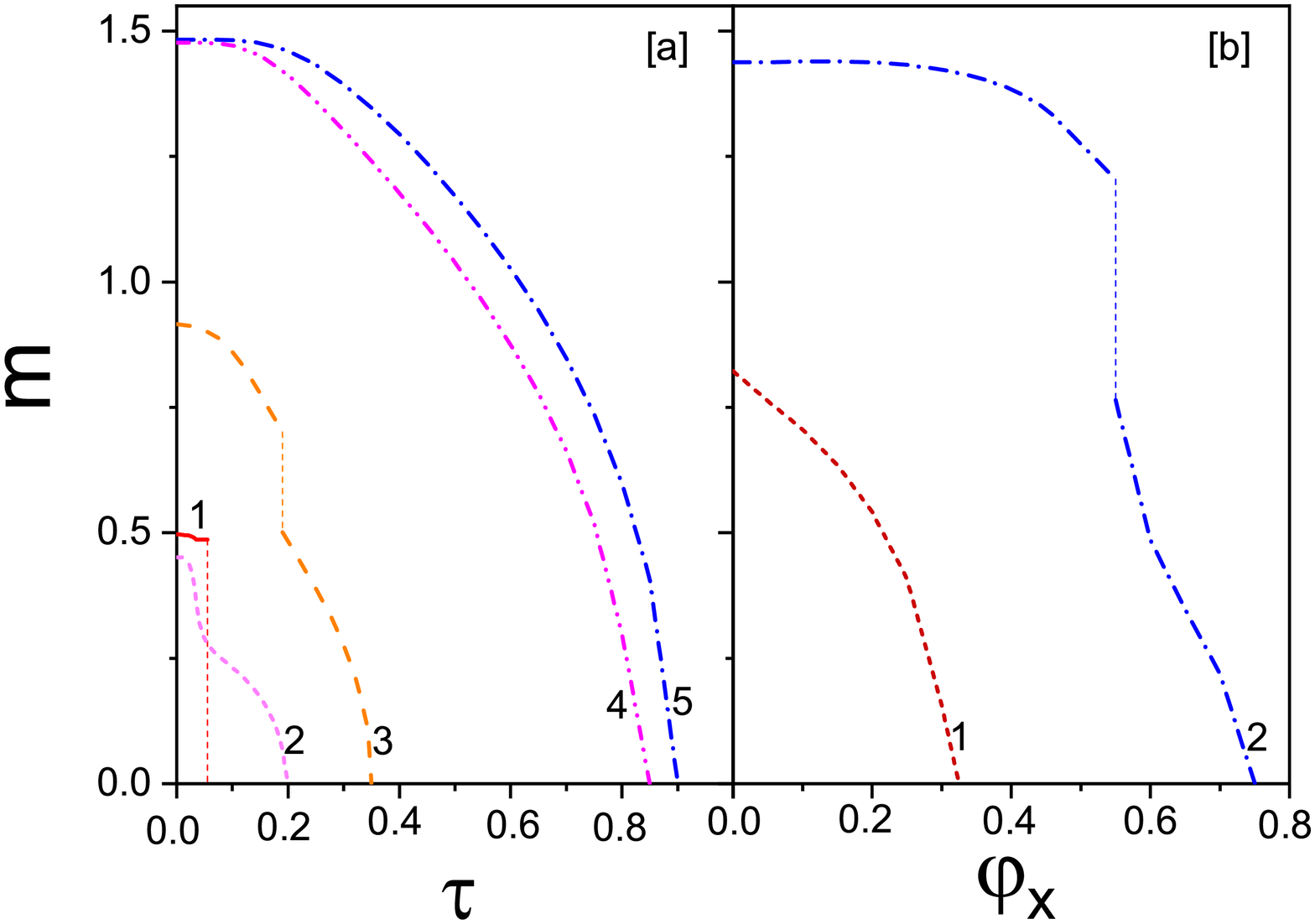}
\caption{(Color online) Behavior of magnetization $m$ versus temperature $\tau$ (a) and parameter $\varphi_{x}$ (b) of the spin-3/2 BC model with two RTSIAs for $q=0.8$ and $\varphi_{y}=0.5$. Here, we have used the following set of parameters: $(a)$ - (1) $p =1.00$ and $\varphi_{x}=0.95$, (2) $p =0.65$ and $\varphi_{x}=1.00$, (3) $p =0.64$ and $\varphi_{x}=2.50$, (4) $p =1.00$ and $\varphi_{x}=0.25$, (5) $p =0.65$ and $\varphi_{x}=0.15$. $(b)$ - (1) $p=1.00$ and $\tau =0.85$, (2) $p=1.00$ and $\tau=0.10$.}
\label{4}
\end{figure}

To analyze the effects of the anisotropy $\varphi_{y}$, which was made equal to zero in the previous analyzes, in Fig.~\ref{4}, we have shown our results for magnetization $m$ as a function of temperature $\tau$ and $\varphi_{x}$ for $\varphi_{y}=0.5$ and $q = 0.8$.  For this case, and others that we have investigated, we observe TCPs (see Fig.~\ref{1}). We can observe in Fig.~\ref{4}(a) in curves (4) (with $p=1.00$ and $\varphi_{x} = 0.25$)  and  (5) (with $p=0.65$ and $\varphi_{x} = 0.15$)    that the magnetization goes to zero continuously, confirming the second-order line that we have observed in Fig.~\ref{1}(d). When we increase the magnitude of $\varphi_x$, we can perceive that the magnetization exhibits a discontinuity, which characterizes the first-order lines as showed in the curves (1) with $p =1.00$ and $\varphi_{x}=0.95$, (2) $p =0.65$ and $\varphi_{x}=1.00$, and (3) $p =0.64$ and $\varphi_{x}=2.50$. This same behavior can be also confirmed in the presented results for the magnetization $m$ as a function of the anisotropy $\varphi_{x}$, which are shown in Fig.~\ref{4}(b), with the fixed temperature. For $\tau = 0.85$ and $p = 1.00$, where all the spins are under the influence of $\varphi_{x}$ (see curve (1)), the magnetization does not present discontinuity. As we displace to regions of greater $\varphi_{x}$ values, the magnetization presents a discontinuity in its shape, which confirm the presence of the first-order line (see  curve (2) with $p=1.00$ $\tau=0.10$ ), therefore, corroborating the topology of the phase diagram of Fig.~\ref{1}(d) .

\begin{figure}[h]
\centering
\includegraphics[scale=0.38]{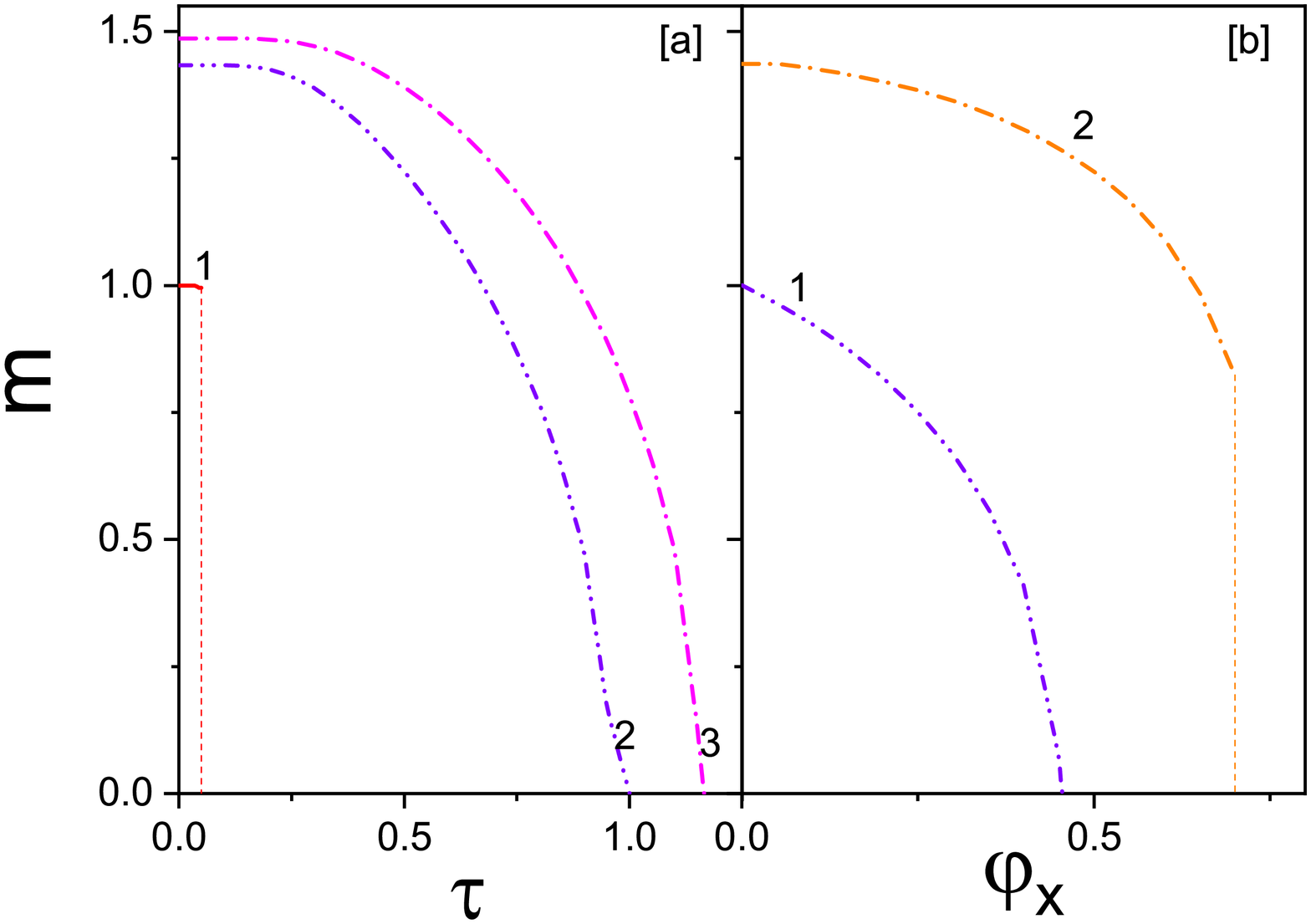}
\caption{(Color online) Behavior of magnetization $m$ versus temperature $\tau$ (a) and parameter $\varphi_{x}$ (b) of the spin-3/2 BC model with two RTSIAs for  $\varphi_{y}=-0.1$. Here, we have used the following set of parameters: $(a)$ $q=0.315$ for the curves (1) $p =1.0$ and $\varphi_x=1.0$ , (2) $p =1.0$ and $\varphi_{x}=0.50$, (3) $p=0.90$ and $\varphi_{x}=0.90$.  $(b)$ $q=0.315$ for the curves (1) $p=1.00$ and $\tau =1.0$, (2) $p=1.00$ and $\tau=0.5$.}
\label{5}
\end{figure}

Finally, we displayed in Fig.~\ref{5} ours results for the behavior of the magnetization $m$ as a function of temperature $\tau$ (Fig.~\ref{5}(a)) with $q=0.315$ and single-ion anisotropy $\varphi_{x}$ (Fig.~\ref{5}(b)) with $q = 0.315$, both for $\varphi_{y}=-0.10$.  In Fig.~\ref{5}(a), the curve (1) for $p =1.00$ and $\varphi_x=1.00$, we can observe the presence of the first-order transition lines, evidenced by the abrupt change in the magnetization behavior. As $\varphi_{x}$ decreases (see curves (2) $p =1.00$ and $\varphi_{x}=0.50$ and (3) $p=0.90$ and $\varphi_{x}=0.90$) the discontinuity disappears and we begin to observe that magnetization goes to zero continuously, which characterizes the second-order transition lines. The magnetization $m$ as a function of $\varphi_{x}$, with fixed temperature is shown in Fig.~\ref{5}(b). For the two cases illustrated, we can see that magnetization presents a typical behavior of second- and first-order transition in curves (1) $p=1.00$ and $\tau =1.00$ and (2) $p=1.00$ and $\tau=0.5$,  thus confirming the topology of the phase diagram of Fig.~\ref{2}(a).

\section{Conclusions}

Our aim in this paper was to study the critical behavior of the spin$-3/2$ BC quantum model on the presence of two RTSIAs by using the variational method based on the Bogoliubov inequality. The RTSIAs are also governed by  probability distribution functions of the type bimodal. We have shown that anisotropy changes the topology of the phase diagram topology concerning to pure model. For some anisotropy values, we have found the presence of tricritical points, which separate second- and first-order transition lines. 

For the case where the magnitude of $\varphi_{y}$ is taken to be equal to zero $q=0$, then  the spins are influenced only by the anisotropy $\varphi_{x}$, we have shown that the presence of TCP only emerges for $p \geq 0.95$,  this is, $95\%$ of the spins are influenced by this anisotropy. In this same Fig.~\ref{1}(a), we have illustrated the pure case ($p = 0$) where we have obtained the critical temperature of $\tau =1.25$. This case is in agreement with the reference~\cite{erhan}. The phase diagrams in  $\tau - \varphi_{x}$ plane for $\varphi_{y}=0.50$ (Figs.~\ref{1}(a), (b), (c), and (d)) have shown that for all the cases investigated we found tricritical behavior. We have seen that the greater the number of spins under the influence of $\varphi_{y}$, that is, the greater values of $q$, the tricritical points occur for values smaller of $p$, this is, a smaller amount of spins under influence of anisotropy $\varphi_{x}$. Our analysis of  Fig.~\ref{1} and  Fig.~\ref{2} have shown that the tricritical behavior is observed in the range of $0 > \varphi_{y} \le -10 $, where for $\varphi_{y}= -0.10$ only for $q \le 0.315$. When we increase the magnitude of this anisotropy to the negative side  $ \varphi_{y}= -0.20$, the first-order transition lines are no longer observed. All these results were confirmed by the magnetization analyzes investigated as a function of temperature $\tau$ and $\varphi_{x}$.

\begin{acknowledgments}

The authors acknowledge financial support from the Brazilian agencies CNPq and CAPES.

\end{acknowledgments}

\appendix
\section{ Free energy Landau-type expansion coefficients}

In order to obtain information about the phase transitions displayed by the system, in the presence of anisotropy, we have used Landau theory, which consists of expanding the free energy in powers of the order parameter, $m$, as given in Eq. (19). In this equation, the coefficients $a_2$, $a_4$ depend on the model parameters, namely:  $\tau$, $\delta_{x}$, $\delta_{y}$, $p$, $q$. They can be obtained using the equations below:

\begin{eqnarray}
 a_{p}=\frac{1}{p!}{\left(\frac{\partial^{p}\Psi}{\partial m_{p}}\right)}_{m=0},
\end{eqnarray}

that provides

\begin{eqnarray}
a_{2}&=&-\frac{1}{2\tau}
+\frac{1}{8e^{\frac{5(\varphi_{x}+\varphi_{y})}{4\tau}}\cosh{\left(\frac{1}{\tau}\sqrt{\varphi_{x}^{2}-\varphi_{x}\varphi_{y}+\varphi_{y}^{2}}\right)}}
.\left[pq\left[e^{\frac{5(\varphi_{x}+\varphi_{y})}{4\tau}}\cosh{\left(\frac{1}{\tau}\sqrt{\varphi_{x}^{2}-\varphi_{x}\varphi_{y}+\varphi_{y}^{2}}\right)}{\tau^{-2}}\right.\right. \nonumber \\
&+&\left.\left.\frac{e^{\frac{5(\varphi_{x}+\varphi_{y})}{4\tau}}\sinh{\left(\frac{1}{\tau}\sqrt{\varphi_{x}^{2}-\varphi_{x}\varphi_{y}+\varphi_{y}^{2}}\right)(-\varphi_{x}-\varphi_{y})}}{\tau^{2}\sqrt{\varphi_{x}^{2}-\varphi_{x}\varphi_{y}+\varphi_{y}^{2}}} \right.\right.
+\left.\left.\frac{e^{\frac{5(\varphi_{x}+\varphi_{y})}{4\tau}}\cosh{\left(\frac{1}{\tau}\sqrt{\varphi_{x}^{2}-\varphi_{x}\varphi_{y}+\varphi_{y}^{2}}\right)(-\varphi_{x}-\varphi_{y})^{2}}}{2\tau^{2}{(\varphi_{x}^{2}-\varphi_{x}\varphi_{y}+\varphi_{y}^{2})}} \right.\right. \nonumber \\
&-&\left.\left.\frac{e^{\frac{5(\varphi_{x}+\varphi_{y})}{4\tau}}\sinh{\left(\frac{1}{\tau}\sqrt{\varphi_{x}^{2}-\varphi_{x}\varphi_{y}+\varphi_{y}^{2}}\right)(-\varphi_{x}-\varphi_{y})^{2}}}{2\tau{(\varphi_{x}^{2}-\varphi_{x}\varphi_{y}+\varphi_{y}^{2})^{\frac{3}{2}}}} \right.\right.
+\left.\left.\frac{4e^{\frac{5(\varphi_{x}+\varphi_{y})}{4\tau}}\sinh{\left(\frac{1}{\tau}\sqrt{\varphi_{x}^{2}-\varphi_{x}\varphi_{y}+\varphi_{y}^{2}}\right)}}{\tau\sqrt{\varphi_{x}^{2}-\varphi_{x}\varphi_{y}+\varphi_{y}^{2}}} \right.\right. \nonumber \\
&-&\left.\left.\frac{e^{\frac{5(\varphi_{x}+\varphi_{y})}{4\tau}}\sinh{\left(\frac{1}{\tau}\sqrt{\varphi_{x}^{2}-\varphi_{x}\varphi_{y}+\varphi_{y}^{2}}\right)(\varphi_{x}+\varphi_{y})}}{\tau^{2}\sqrt{\varphi_{x}^{2}-\varphi_{x}\varphi_{y}+\varphi_{y}^{2}}} \right.\right. +\left.\left.\frac{e^{\frac{5(\varphi_{x}+\varphi_{y})}{4\tau}}\cosh{\left(\frac{1}{\tau}\sqrt{\varphi_{x}^{2}-\varphi_{x}\varphi_{y}+\varphi_{y}^{2}}\right)(\varphi_{x}+\varphi_{y})^{2}}}{2\tau^{2}{(\varphi_{x}^{2}-\varphi_{x}\varphi_{y}+\varphi_{y}^{2})}}\right.\right. \nonumber \\
&-&\left.\left.\frac{e^{\frac{5(\varphi_{x}+\varphi_{y})}{4\tau}}\sinh{\left(\frac{1}{\tau}\sqrt{\varphi_{x}^{2}-\varphi_{x}\varphi_{y}+\varphi_{y}^{2}}\right)(\varphi_{x}+\varphi_{y})^{2}}}{2\tau{(\varphi_{x}^{2}-\varphi_{x}\varphi_{y}+\varphi_{y}^{2})^{\frac{3}{2}}}}\right]\right]
-\frac{1}{32\left(e^{\frac{5(\varphi_{x}+\varphi_{y})}{4\tau}}\right)^{2}\cosh{\left(\frac{1}{\tau}\sqrt{\varphi_{x}^{2}-\varphi_{x}\varphi_{y}+\varphi_{y}^{2}}\right)^{2}}} \nonumber \\
&.&\left[pq\left[\frac{e^{\frac{5(\varphi_{x}+\varphi_{y})}{4\tau}}\sinh{\left(\frac{1}{\tau}\sqrt{\varphi_{x}^{2}-\varphi_{x}\varphi_{y}+\varphi_{y}^{2}}\right)(-\varphi_{x}-\varphi_{y})}}{\tau\sqrt{\varphi_{x}^{2}-\varphi_{x}\varphi_{y}+\varphi_{y}^{2}}}\right.\right.
+\left.\left.\frac{e^{\frac{5(\varphi_{x}+\varphi_{y})}{4\tau}}\sinh{\left(\frac{1}{\tau}\sqrt{\varphi_{x}^{2}-\varphi_{x}\varphi_{y}+\varphi_{y}^{2}}\right)(\varphi_{x}+\varphi_{y})}}{\tau\sqrt{\varphi_{x}^{2}-\varphi_{x}\varphi_{y}+\varphi_{y}^{2}}}\right]^{2}\right] \nonumber \\
&+&\frac{1}{8e^{\frac{5\varphi_{x}}{4\tau}}\cosh{\left(\frac{\varphi_{x}}{\tau}\right)} }
.\left[p(1-q)\left[\frac{2e^{\frac{5\varphi_{x}}{4\tau}}\cosh{\left(\frac{\varphi_{x}}{\tau}\right)}}{\tau^{2}} \right.\right.
-\left.\left.\frac{2e^{\frac{5\varphi_{x}}{4\tau}}\sinh{\left(\frac{\varphi_{x}}{\tau}\right)}}{\tau^{2}} \right.\right.
+\left.\left.\frac{3e^{\frac{5\varphi_{x}}{4\tau}}\sinh{\left(\frac{\varphi_{x}}{\tau}\right)}}{\tau\varphi_{x}}\right]\right] \nonumber \\
&+&\frac{1}{8e^{\frac{5\varphi_{y}}{4\tau}}\cosh{\left(\frac{\varphi_{y}}{\tau}\right)}}
.\left[q(1-p)\left[\frac{2e^{\frac{5\varphi_{y}}{4\tau}}\cosh{\left(\frac{\varphi_{y}}{\tau}\right)}}{\tau^{2}}\right.\right.
-\left.\left.\frac{2e^{\frac{5\varphi_{y}}{4\tau}}\sinh{\left(\frac{\varphi_{y}}{\tau}\right)}}{\tau^{2}}\right.\right.
+\left.\left.\frac{3e^{\frac{5(\varphi_{y})}{4\tau}}\sinh{\left(\frac{\varphi_{y}}{\tau}\right)}}{\tau\varphi_{y}}\right]\right] \nonumber \\
&+&\frac{5(1-p)(1-q)}{8\tau^{2}},
\end{eqnarray}
and
\begin{eqnarray}
a_{4}&=&-\frac{1}{96\left(e^{\frac{5(\varphi_{x}+\varphi_{y})}{4\tau}}\right)^{2}\cosh{\left(\frac{1}{\tau}\sqrt{\varphi_{x}^{2}-\varphi_{x}\varphi_{y}+\varphi_{y}^{2}}\right)^{2}}}
.\left[pq\left[\frac{3e^{\frac{5(\varphi_{x}+\varphi_{y})}{4\tau}}\sinh{\left(\frac{1}{\tau}\sqrt{\varphi_{x}^{2}-\varphi_{x}\varphi_{y}+\varphi_{y}^{2}}\right)(-\varphi_{x}-\varphi_{y})^{2}}}{4\tau^{2}{(\varphi_{x}^{2}-\varphi_{x}\varphi_{y}+\varphi_{y}^{2})^{\frac{3}{2}}}}\right.\right. \nonumber \\
&-&\left.\left.\frac{3e^{\frac{5(\varphi_{x}+\varphi_{y})}{4\tau}}\cosh{\left(\frac{1}{\tau}\sqrt{\varphi_{x}^{2}-\varphi_{x}\varphi_{y}+\varphi_{y}^{2}}\right)(-\varphi_{x}-\varphi_{y})^{3}}}{4\tau^{2}{(\varphi_{x}^{2}-\varphi_{x}\varphi_{y}+\varphi_{y}^{2})^{2}}}\right.\right.
+\left.\left.\frac{3e^{\frac{5(\varphi_{x}+\varphi_{y})}{4\tau^{2}}}\sinh{\left(\frac{1}{\tau}\sqrt{\varphi_{x}^{2}-\varphi_{x}\varphi_{y}+\varphi_{y}^{2}}\right)(-\varphi_{x}-\varphi_{y})^{3}}}{4\tau{(\varphi_{x}^{2}-\varphi_{x}\varphi_{y}+\varphi_{y}^{2})^{\frac{3}{2}}}}\right.\right. \nonumber \\
&+&\left.\left.\frac{3e^{\frac{5(\varphi_{x}+\varphi_{y})}{4\tau}}\sinh{\left(\frac{1}{\tau}\sqrt{\varphi_{x}^{2}-\varphi_{x}\varphi_{y}+\varphi_{y}^{2}}\right)(\varphi_{x}+\varphi_{y})^{2}}}{4\tau^{2}{(\varphi_{x}^{2}-\varphi_{x}\varphi_{y}+\varphi_{y}^{2})^{\frac{3}{2}}}} \right.\right.
-\left.\left.\frac{3e^{\frac{5(\varphi_{x}+\varphi_{y})}{4\tau}}\cosh{\left(\frac{1}{\tau}\sqrt{\varphi_{x}^{2}-\varphi_{x}\varphi_{y}+\varphi_{y}^{2}}\right)(\varphi_{x}+\varphi_{y})^{3}}}{4\tau^{2}{(\varphi_{x}^{2}-\varphi_{x}\varphi_{y}+\varphi_{y}^{2})^{2}}} \right.\right. \nonumber \\
&+&\left.\left.\frac{3e^{\frac{5(\varphi_{x}+\varphi_{y})}{4\tau}}\sinh{\left(\frac{1}{\tau}\sqrt{\varphi_{x}^{2}-\varphi_{x}\varphi_{y}+\varphi_{y}^{2}}\right)(\varphi_{x}+\varphi_{y})^{3}}}{4\tau{(\varphi_{x}^{2}-\varphi_{x}\varphi_{y}+\varphi_{y}^{2})^{\frac{3}{2}}}} \right.\right.
+\left.\left.\frac{3e^{\frac{5(\varphi_{x}+\varphi_{y})}{4\tau}}\sinh{\left(\frac{1}{\tau}\sqrt{\varphi_{x}^{2}-\varphi_{x}\varphi_{y}+\varphi_{y}^{2}}\right)(\varphi_{x}+\varphi_{y})}}{4\tau^{3}{\sqrt{\varphi_{x}^{2}-\varphi_{x}\varphi_{y}+\varphi_{y}^{2}}}} \right.\right. \nonumber \\
&-&\left.\left.\frac{3e^{\frac{5(\varphi_{x}+\varphi_{y})}{4\tau}}\cosh{\left(\frac{1}{\tau}\sqrt{\varphi_{x}^{2}-\varphi_{x}\varphi_{y}+\varphi_{y}^{2}}\right)(\varphi_{x}+\varphi_{y})^{2}}}{4\tau^{3}{(\varphi_{x}^{2}-\varphi_{x}\varphi_{y}+\varphi_{y}^{2})}} \right.\right.
+\left.\left.\frac{1e^{\frac{5(\varphi_{x}+\varphi_{y})}{4\tau}}\sinh{\left(\frac{1}{\tau}\sqrt{\varphi_{x}^{2}-\varphi_{x}\varphi_{y}+\varphi_{y}^{2}}\right)(\varphi_{x}+\varphi_{y})^{3}}}{4\tau^{3}{(\varphi_{x}^{2}-\varphi_{x}\varphi_{y}+\varphi_{y}^{2})^{\frac{3}{2}}}}  \right.\right. \nonumber \\
&+&\left.\left.\frac{3e^{\frac{5(\varphi_{x}+\varphi_{y})}{4\tau}}\cosh{\left(\frac{1}{\tau}\sqrt{\varphi_{x}^{2}-\varphi_{x}\varphi_{y}+\varphi_{y}^{2}}\right)(\varphi_{x}+\varphi_{y})}}{\tau^{2}{(\varphi_{x}^{2}-\varphi_{x}\varphi_{y}+\varphi_{y}^{2})}} \right.\right.
-\left.\left.\frac{3e^{\frac{5(\varphi_{x}+\varphi_{y})}{4\tau}}\sinh{\left(\frac{1}{\tau}\sqrt{\varphi_{x}^{2}-\varphi_{x}\varphi_{y}+\varphi_{y}^{2}}\right)(\varphi_{x}+\varphi_{y})}}{\tau{(\varphi_{x}^{2}-\varphi_{x}\varphi_{y}+\varphi_{y}^{2})^{\frac{3}{2}}}} \right.\right. \nonumber \\
&+&\left.\left.\frac{3e^{\frac{5(\varphi_{x}+\varphi_{y})}{4\tau}}\sinh{\left(\frac{1}{\tau}\sqrt{\varphi_{x}^{2}-\varphi_{x}\varphi_{y}+\varphi_{y}^{2}}\right)(-\varphi_{x}-\varphi_{y})}}{4\tau^{3}{\sqrt{\varphi_{x}^{2}-\varphi_{x}\varphi_{y}+\varphi_{y}^{2}}}} \right.\right.
+\left.\left.\frac{3e^{\frac{5(\varphi_{x}+\varphi_{y})}{4\tau}}\cosh{\left(\frac{1}{\tau}\sqrt{\varphi_{x}^{2}-\varphi_{x}\varphi_{y}+\varphi_{y}^{2}}\right)(-\varphi_{x}-\varphi_{y})^{2}}}{4\tau^{3}{(\varphi_{x}^{2}-\varphi_{x}\varphi_{y}+\varphi_{y}^{2})}} \right.\right. \nonumber \\
&+&\left.\left.\frac{1e^{\frac{5(\varphi_{x}+\varphi_{y})}{4\tau}}\sinh{\left(\frac{1}{\tau}\sqrt{\varphi_{x}^{2}-\varphi_{x}\varphi_{y}+\varphi_{y}^{2}}\right)(-\varphi_{x}-\varphi_{y})^{3}}}{4\tau^{3}{(\varphi_{x}^{2}-\varphi_{x}\varphi_{y}+\varphi_{y}^{2})^{\frac{3}{2}}}} \right.\right.
+\left.\left.\frac{3e^{\frac{5(\varphi_{x}+\varphi_{y})}{4\tau}}\cosh{\left(\frac{1}{\tau}\sqrt{\varphi_{x}^{2}-\varphi_{x}\varphi_{y}+\varphi_{y}^{2}}\right)(-\varphi_{x}-\varphi_{y})}}{\tau^{2}{(\varphi_{x}^{2}-\varphi_{x}\varphi_{y}+\varphi_{y}^{2})}} \right.\right. \nonumber \\
&-&\left.\left.\frac{3e^{\frac{5(\varphi_{x}+\varphi_{y})}{4\tau}}\sinh{\left(\frac{1}{\tau}\sqrt{\varphi_{x}^{2}-\varphi_{x}\varphi_{y}+\varphi_{y}^{2}}\right)(-\varphi_{x}-\varphi_{y})}}{\tau{(\varphi_{x}^{2}-\varphi_{x}\varphi_{y}+\varphi_{y}^{2})^{\frac{3}{2}}}}\right] \right.
.\left.\left[\frac{e^{\frac{5(\varphi_{x}+\varphi_{y})}{4\tau}}\sinh{\left(\frac{1}{\tau}\sqrt{\varphi_{x}^{2}-\varphi_{x}\varphi_{y}+\varphi_{y}^{2}}\right)(-\varphi_{x}-\varphi_{y})^{3}}}{\tau{\sqrt{\varphi_{x}^{2}-\varphi_{x}\varphi_{y}+\varphi_{y}^{2}}}} \right.\right. \nonumber \\
&+&\left.\left.\frac{e^{\frac{5(\varphi_{x}+\varphi_{y})}{4\tau}}\sinh{\left(\frac{1}{\tau}\sqrt{\varphi_{x}^{2}-\varphi_{x}\varphi_{y}+\varphi_{y}^{2}}\right)(\varphi_{x}+\varphi_{y})}}{\tau{\sqrt{\varphi_{x}^{2}-\varphi_{x}\varphi_{y}+\varphi_{y}^{2}}}}\right]\right]
-\frac{1}{128\left(e^{\frac{5(\varphi_{x}+\varphi_{y})}{4\tau}}\right)^{2}\cosh{\left(\frac{1}{\tau}\sqrt{\varphi_{x}^{2}-\varphi_{x}\varphi_{y}+\varphi_{y}^{2}}\right)^{2}}} \nonumber \\
&.&\left[pq\left[\frac{e^{\frac{5(\varphi_{x}+\varphi_{y})}{4\tau}}\cosh{\left(\frac{1}{\tau}\sqrt{\varphi_{x}^{2}-\varphi_{x}\varphi_{y}+\varphi_{y}^{2}}\right)}}{\tau^{2}} \right.\right.
+\left.\left.\frac{e^{\frac{5(\varphi_{x}+\varphi_{y})}{4\tau}}\sinh{\left(\frac{1}{\tau}\sqrt{\varphi_{x}^{2}-\varphi_{x}\varphi_{y}+\varphi_{y}^{2}}\right)(-\varphi_{x}-\varphi_{y})}}{\tau^{2}\sqrt{\varphi_{x}^{2}-\varphi_{x}\varphi_{y}+\varphi_{y}^{2}}} \right.\right.\nonumber \\
&+&\left.\left.\frac{e^{\frac{5(\varphi_{x}+\varphi_{y})}{4\tau}}\cosh{\left(\frac{1}{\tau}\sqrt{\varphi_{x}^{2}-\varphi_{x}\varphi_{y}+\varphi_{y}^{2}}\right)(-\varphi_{x}-\varphi_{y})^{2}}}{2\tau^{2}{(\varphi_{x}^{2}-\varphi_{x}\varphi_{y}+\varphi_{y}^{2})}} \right.\right.
-\left.\left.\frac{e^{\frac{5(\varphi_{x}+\varphi_{y})}{4\tau}}\sinh{\left(\frac{1}{\tau}\sqrt{\varphi_{x}^{2}-\varphi_{x}\varphi_{y}+\varphi_{y}^{2}}\right)(-\varphi_{x}-\varphi_{y})^{2}}}{2\tau{(\varphi_{x}^{2}-\varphi_{x}\varphi_{y}+\varphi_{y}^{2})^{\frac{3}{2}}}} \right.\right.\nonumber \\
&+&\left.\left.\frac{4e^{\frac{5(\varphi_{x}+\varphi_{y})}{4\tau}}\sinh{\left(\frac{1}{\tau}\sqrt{\varphi_{x}^{2}-\varphi_{x}\varphi_{y}+\varphi_{y}^{2}}\right)}}{\tau\sqrt{\varphi_{x}^{2}-\varphi_{x}\varphi_{y}+\varphi_{y}^{2}}} \right.\right.
-\left.\left.\frac{e^{\frac{5(\varphi_{x}+\varphi_{y})}{4\tau}}\sinh{\left(\frac{1}{\tau}\sqrt{\varphi_{x}^{2}-\varphi_{x}\varphi_{y}+\varphi_{y}^{2}}\right)(\varphi_{x}+\varphi_{y})}}{\tau^{2}\sqrt{\varphi_{x}^{2}-\varphi_{x}\varphi_{y}+\varphi_{y}^{2}}} \right.\right.\nonumber \\
&+&\left.\left.\frac{e^{\frac{5(\varphi_{x}+\varphi_{y})}{4\tau}}\cosh{\left(\frac{1}{\tau}\sqrt{\varphi_{x}^{2}-\varphi_{x}\varphi_{y}+\varphi_{y}^{2}}\right)(\varphi_{x}+\varphi_{y})^{2}}}{2\tau^{2}{(\varphi_{x}^{2}-\varphi_{x}\varphi_{y}+\varphi_{y}^{2})}} \right.\right.
-\left.\left.\frac{e^{\frac{5(\varphi_{x}+\varphi_{y})}{4\tau}}\sinh{\left(\frac{1}{\tau}\sqrt{\varphi_{x}^{2}-\varphi_{x}\varphi_{y}+\varphi_{y}^{2}}\right)(\varphi_{x}+\varphi_{y})^{2}}}{2\tau{(\varphi_{x}^{2}-\varphi_{x}\varphi_{y}+\varphi_{y}^{2})^{\frac{3}{2}}}}\right]^{2}\right]  \nonumber \\
&-&\frac{1}{1024\left(e^{\frac{5(\varphi_{x}+\varphi_{y})}{4\tau}}\right)^{4}\cosh{\left(\frac{1}{\tau}\sqrt{\varphi_{x}^{2}-\varphi_{x}\varphi_{y}+\varphi_{y}^{2}}\right)^{4}}}
.\left[pq\left[\frac{e^{\frac{5(\varphi_{x}+\varphi_{y})}{4\tau}}\sinh{\left(\frac{1}{\tau}\sqrt{\varphi_{x}^{2}-\varphi_{x}\varphi_{y}+\varphi_{y}^{2}}\right)(-\varphi_{x}-\varphi_{y})}}{\tau\sqrt{\varphi_{x}^{2}-\varphi_{x}\varphi_{y}+\varphi_{y}^{2}}} \right.\right.\nonumber \\
&+&\left.\left.\frac{e^{\frac{5(\varphi_{x}+\varphi_{y})}{4\tau}}\sinh{\left(\frac{1}{\tau}\sqrt{\varphi_{x}^{2}-\varphi_{x}\varphi_{y}+\varphi_{y}^{2}}\right)(\varphi_{x}+\varphi_{y})}}{\tau\sqrt{\varphi_{x}^{2}-\varphi_{x}\varphi_{y}+\varphi_{y}^{2}}}\right]^{4}\right]
\left.\left.-\frac{1}{128\left(e^{\frac{5\varphi_{x}}{4\tau}}\right)^{2}\cosh{\left(\frac{\varphi_{x}}{\tau}\right)^{2}}} \right.\right.
.\left[p(1-q)\left[\frac{2e^{\frac{5\varphi_{x}}{4\tau}}\cosh{\left(\frac{\varphi_{x}}{\tau}\right)}}{\tau^{2}} \right.\right.\nonumber \\
&-&\left.\left.\frac{2e^{\frac{5\varphi_{x}}{4\tau}}\sinh{\left(\frac{\varphi_{x}}{\tau}\right)}}{\tau^{2}} \right.\right.
+\left.\left.\frac{3e^{\frac{5\varphi_{x}}{4\tau}}\sinh{\left(\frac{\varphi_{x}}{\tau}\right)}}{\tau\varphi_{x}}\right]^{2}\right]
-\frac{1}{128\left(e^{\frac{5\varphi_{y}}{4\tau}}\right)^{2}\cosh{\left(\frac{\varphi_{y}}{\tau}\right)^{2}}}
.\left[q(1-p)\left[\frac{2e^{\frac{5\varphi_{y}}{4\tau}}\cosh{\left(\frac{\varphi_{y}}{\tau}\right)}}{\tau^{2}} \right.\right. \nonumber \\
&-&\left.\left.\frac{2e^{\frac{5\varphi_{y}}{4\tau}}\sinh{\left(\frac{\varphi_{y}}{\tau}\right)}}{\tau^{2}} \right.\right.
+\left.\left.\frac{3e^{\frac{5\varphi_{y}}{4\tau}}\sinh{\left(\frac{\varphi_{y}}{\tau}\right)}}{\tau\varphi_{y}}\right]^{2}\right]
-\frac{17(1-p)(1-q)}{192\tau^{2}}
-\frac{1}{128\left(e^{\frac{5(\varphi_{x}+\varphi_{y})}{4\tau}}\right)^{3}\cosh{\left(\frac{1}{\tau}\sqrt{\varphi_{x}^{2}-\varphi_{x}\varphi_{y}+\varphi_{y}^{2}}\right)^{3}}} \nonumber \\
&.&\left[pq\left[\frac{e^{\frac{5(\varphi_{x}+\varphi_{y})}{4\tau}}\cosh{\left(\frac{1}{\tau}\sqrt{\varphi_{x}^{2}-\varphi_{x}\varphi_{y}+\varphi_{y}^{2}}\right)}}{\tau^{2}} \right.\right.
+\left.\left.\frac{e^{\frac{5(\varphi_{x}+\varphi_{y})}{4\tau}}\sinh{\left(\frac{1}{\tau}\sqrt{\varphi_{x}^{2}-\varphi_{x}\varphi_{y}+\varphi_{y}^{2}}\right)(-\varphi_{x}-\varphi_{y})}}{\tau^{2}\sqrt{\varphi_{x}^{2}-\varphi_{x}\varphi_{y}+\varphi_{y}^{2}}} \right.\right. \nonumber \\
&+&\left.\left.\frac{e^{\frac{5(\varphi_{x}+\varphi_{y})}{4\tau}}\cosh{\left(\frac{1}{\tau}\sqrt{\varphi_{x}^{2}-\varphi_{x}\varphi_{y}+\varphi_{y}^{2}}\right)(-\varphi_{x}-\varphi_{y})^{2}}}{2\tau^{2}{(\varphi_{x}^{2}-\varphi_{x}\varphi_{y}+\varphi_{y}^{2})}} \right.\right.
-\left.\left.\frac{e^{\frac{5(\varphi_{x}+\varphi_{y})}{4\tau}}\sinh{\left(\frac{1}{\tau}\sqrt{\varphi_{x}^{2}-\varphi_{x}\varphi_{y}+\varphi_{y}^{2}}\right)(-\varphi_{x}-\varphi_{y})^{2}}}{2\tau{(\varphi_{x}^{2}-\varphi_{x}\varphi_{y}+\varphi_{y}^{2})^{\frac{3}{2}}}} \right.\right. \nonumber \\
&+&\left.\left.\frac{4e^{\frac{5(\varphi_{x}+\varphi_{y})}{4\tau}}\sinh{\left(\frac{1}{\tau}\sqrt{\varphi_{x}^{2}-\varphi_{x}\varphi_{y}+\varphi_{y}^{2}}\right)}}{\tau\sqrt{\varphi_{x}^{2}-\varphi_{x}\varphi_{y}+\varphi_{y}^{2}}} \right.\right.
-\left.\left.\frac{e^{\frac{5(\varphi_{x}+\varphi_{y})}{4\tau}}\sinh{\left(\frac{1}{\tau}\sqrt{\varphi_{x}^{2}-\varphi_{x}\varphi_{y}+\varphi_{y}^{2}}\right)(\varphi_{x}+\varphi_{y})}}{\tau^{2}\sqrt{\varphi_{x}^{2}-\varphi_{x}\varphi_{y}+\varphi_{y}^{2}}} \right.\right. \nonumber \\
&+&\left.\left.\frac{e^{\frac{5(\varphi_{x}+\varphi_{y})}{4\tau}}\cosh{\left(\frac{1}{\tau}\sqrt{\varphi_{x}^{2}-\varphi_{x}\varphi_{y}+\varphi_{y}^{2}}\right)(\varphi_{x}+\varphi_{y})^{2}}}{2\tau^{2}{(\varphi_{x}^{2}-\varphi_{x}\varphi_{y}+\varphi_{y}^{2})}} \right.\right.
-\left.\left.\frac{e^{\frac{5(\varphi_{x}+\varphi_{y})}{4\tau}}\sinh{\left(\frac{1}{\tau}\sqrt{\varphi_{x}^{2}-\varphi_{x}\varphi_{y}+\varphi_{y}^{2}}\right)(\varphi_{x}+\varphi_{y})^{2}}}{2\tau{(\varphi_{x}^{2}-\varphi_{x}\varphi_{y}+\varphi_{y}^{2})^{\frac{3}{2}}}}\right]\right.  \nonumber \\
&.&\left.\left[\frac{e^{\frac{5(\varphi_{x}+\varphi_{y})}{4\tau}}\sinh{\left(\frac{1}{\tau}\sqrt{\varphi_{x}^{2}-\varphi_{x}\varphi_{y}+\varphi_{y}^{2}}\right)(-\varphi_{x}-\varphi_{y})}}{\tau\sqrt{\varphi_{x}^{2}-\varphi_{x}\varphi_{y}+\varphi_{y}^{2}}} \right.\right.
+\left.\left.\frac{e^{\frac{5(\varphi_{x}+\varphi_{y})}{4\tau}}\sinh{\left(\frac{1}{\tau}\sqrt{\varphi_{x}^{2}-\varphi_{x}\varphi_{y}+\varphi_{y}^{2}}\right)(\varphi_{x}+\varphi_{y})}}{\tau\sqrt{\varphi_{x}^{2}-\varphi_{x}\varphi_{y}+\varphi_{y}^{2}}}\right]^{2}\right] \nonumber \\
&+&\frac{1}{96e^{\frac{5\varphi_{x}}{4\tau}}\cosh{\left(\frac{\varphi_{x}}{\tau}\right)}}
.\left[p(1-q)\left[\frac{2e^{\frac{5\varphi_{x}}{4\tau}}\cosh{\left(\frac{\varphi_{x}}{\tau}\right)}}{\tau^{4}} \right.\right.
-\left.\left.\frac{2e^{\frac{5\varphi_{x}}{4\tau}}\sinh{\left(\frac{\varphi_{x}}{\tau}\right)}}{\tau^{4}} \right.\right.
+\left.\left.\frac{25e^{\frac{5\varphi_{x}}{4\tau}}\sinh{\left(\frac{\varphi_{x}}{\tau}\right)}}{4\tau^{2}\varphi_{x}^{2}} \right.\right. \nonumber \\
&+&\left.\left.\frac{9e^{\frac{5\varphi_{x}}{4\tau}}\sinh{\left(\frac{\varphi_{x}}{\tau}\right)}}{4\tau\varphi_{x}^{3}}\right]\right]
+\left.\left.\frac{1}{96e^{\frac{5\varphi_{y}}{4\tau}}\cosh{\left(\frac{\varphi_{y}}{\tau}\right)}} \right.\right.
.\left[q(1-p)\left[\frac{9e^{\frac{5\varphi_{y}}{4\tau}}\sinh{\left(\frac{\varphi_{y}}{\tau}\right)}}{\tau^{3}\varphi_{y}}  \right.\right.
-\left.\left.\frac{9e^{\frac{5\varphi_{y}}{4\tau}}\sinh{\left(\frac{\varphi_{y}}{\tau}\right)}}{4\tau\varphi_{y}^{3}} \right.\right.\nonumber \\
&-&\left.\left.\frac{9e^{\frac{5\varphi_{y}}{4\tau}}\cosh{\left(\frac{\varphi_{y}}{\tau}\right)}}{\tau^{3}\varphi_{y}} \right.\right.
+\left.\left.\frac{2e^{\frac{5\varphi_{y}}{4\tau}}\cosh{\left(\frac{\varphi_{y}}{\tau}\right)}}{\tau^{4}} \right.\right.
-\left.\left.\frac{9e^{\frac{5\varphi_{y}}{4\tau}}\cosh{\left(\frac{\varphi_{y}}{\tau}\right)}}{\tau^{2}\varphi_{y}^{2}} \right.\right.
-\left.\left.\frac{2e^{\frac{5\varphi_{y}}{4\tau}}\sinh{\left(\frac{\varphi_{y}}{\tau}\right)}}{\tau^{4}} \right.\right.   +\left.\left.\frac{9e^{\frac{5\varphi_{y}}{4\tau}}\sinh{\left(\frac{\varphi_{y}}{\tau}\right)}}{\tau^{2}\varphi_{y}^{2}}\right]\right] \nonumber \\
&+&\frac{1}{96e^{\frac{5(\varphi_{x}+\varphi_{y})}{4\tau}}\cosh{\left(\frac{1}{\tau}\sqrt{\varphi_{x}^{2}-\varphi_{x}\varphi_{y}+\varphi_{y}^{2}}\right)}}
.\left[pq\left[\frac{12e^{\frac{5(\varphi_{x}+\varphi_{y})}{4\tau}}\cosh{\left(\frac{1}{\tau}\sqrt{\varphi_{x}^{2}-\varphi_{x}\varphi_{y}+\varphi_{y}^{2}}\right)}}{\tau^{2}(\varphi_{x}^{2}-\varphi_{x}\varphi_{y}+\varphi_{y}^{2})} \right.\right. \nonumber \\
&-&\left.\left.\frac{12e^{\frac{5(\varphi_{x}+\varphi_{y})}{4\tau}}\sinh{\left(\frac{1}{\tau}\sqrt{\varphi_{x}^{2}-\varphi_{x}\varphi_{y}+\varphi_{y}^{2}}\right)}}{\tau(\varphi_{x}^{2}-\varphi_{x}\varphi_{y}+\varphi_{y}^{2})^{\frac{3}{2}}} \right.\right.
+\left.\left.\frac{6e^{\frac{5(\varphi_{x}+\varphi_{y})}{4\tau}}\sinh{\left(\frac{1}{\tau}\sqrt{\varphi_{x}^{2}-\varphi_{x}\varphi_{y}+\varphi_{y}^{2}}\right)}}{\tau^{3}\sqrt{\varphi_{x}^{2}-\varphi_{x}\varphi_{y}+\varphi_{y}^{2}}} \right.\right. \nonumber \\
&+&\left.\left.\frac{e^{\frac{5(\varphi_{x}+\varphi_{y})}{4\tau}}\cosh{\left(\frac{1}{\tau}\sqrt{\varphi_{x}^{2}-\varphi_{x}\varphi_{y}+\varphi_{y}^{2}}\right)}}{4\tau^{4}} \right.\right.
+\left.\left.\frac{e^{\frac{5(\varphi_{x}+\varphi_{y})}{4\tau}}\sinh{\left(\frac{1}{\tau}\sqrt{\varphi_{x}^{2}-\varphi_{x}\varphi_{y}+\varphi_{y}^{2}}\right)}(-\varphi_{x}-\varphi_{y})}{2\tau^{4}\sqrt{\varphi_{x}^{2}-\varphi_{x}\varphi_{y}+\varphi_{y}^{2}}} \right.\right. \nonumber \\
&-&\left.\left.\frac{e^{\frac{5(\varphi_{x}+\varphi_{y})}{4\tau}}\sinh{\left(\frac{1}{\tau}\sqrt{\varphi_{x}^{2}-\varphi_{x}\varphi_{y}+\varphi_{y}^{2}}\right)}(\varphi_{x}+\varphi_{y})}{2\tau^{4}\sqrt{\varphi_{x}^{2}-\varphi_{x}\varphi_{y}+\varphi_{y}^{2}}} \right.\right.
+\left.\left.\frac{3e^{\frac{5(\varphi_{x}+\varphi_{y})}{4\tau}}\cosh{\left(\frac{1}{\tau}\sqrt{\varphi_{x}^{2}-\varphi_{x}\varphi_{y}+\varphi_{y}^{2}}\right)}(\varphi_{x}+\varphi_{y})^{2}}{4\tau^{4}(\varphi_{x}^{2}-\varphi_{x}\varphi_{y}+\varphi_{y}^{2})} \right.\right.\nonumber \\
&-&\left.\left.\frac{e^{\frac{5(\varphi_{x}+\varphi_{y})}{4\tau}}\sinh{\left(\frac{1}{\tau}\sqrt{\varphi_{x}^{2}-\varphi_{x}\varphi_{y}+\varphi_{y}^{2}}\right)}(\varphi_{x}+\varphi_{y})^{3}}{2\tau^{4}(\varphi_{x}^{2}-\varphi_{x}\varphi_{y}+\varphi_{y}^{2})^{\frac{3}{2}}} \right.\right.
-\left.\left.\frac{6e^{\frac{5(\varphi_{x}+\varphi_{y})}{4\tau}}\cosh{\left(\frac{1}{\tau}\sqrt{\varphi_{x}^{2}-\varphi_{x}\varphi_{y}+\varphi_{y}^{2}}\right)}(\varphi_{x}+\varphi_{y})}{\tau^{3}(\varphi_{x}^{2}-\varphi_{x}\varphi_{y}+\varphi_{y}^{2})} \right.\right.\nonumber \\
&+&\left.\left.\frac{e^{\frac{5(\varphi_{x}+\varphi_{y})}{4\tau}}\cosh{\left(\frac{1}{\tau}\sqrt{\varphi_{x}^{2}-\varphi_{x}\varphi_{y}+\varphi_{y}^{2}}\right)}(\varphi_{x}+\varphi_{y})^{4}}{8\tau^{4}(\varphi_{x}^{2}-\varphi_{x}\varphi_{y}+\varphi_{y}^{2})} \right.\right.
+\left.\left.\frac{3e^{\frac{5(\varphi_{x}+\varphi_{y})}{4\tau}}\cosh{\left(\frac{1}{\tau}\sqrt{\varphi_{x}^{2}-\varphi_{x}\varphi_{y}+\varphi_{y}^{2}}\right)}(-\varphi_{x}-\varphi_{y})^{2}}{4\tau^{4}(\varphi_{x}^{2}-\varphi_{x}\varphi_{y}+\varphi_{y})} \right.\right. \nonumber \\
&+&\left.\left.\frac{e^{\frac{5(\varphi_{x}+\varphi_{y})}{4\tau}}\sinh{\left(\frac{1}{\tau}\sqrt{\varphi_{x}^{2}-\varphi_{x}\varphi_{y}+\varphi_{y}^{2}}\right)}(-\varphi_{x}-\varphi_{y})^{3}}{2\tau^{4}(\varphi_{x}^{2}-\varphi_{x}\varphi_{y}+\varphi_{y}^{2})^{\frac{3}{2}}} \right.\right.
+\left.\left.\frac{6e^{\frac{5(\varphi_{x}+\varphi_{y})}{4\tau}}\cosh{\left(\frac{1}{\tau}\sqrt{\varphi_{x}^{2}-\varphi_{x}\varphi_{y}+\varphi_{y}^{2}}\right)}(-\varphi_{x}-\varphi_{y})}{\tau^{3}(\varphi_{x}^{2}-\varphi_{x}\varphi_{y}+\varphi_{y})} \right.\right.\nonumber \\
&+&\left.\left.\frac{e^{\frac{5(\varphi_{x}+\varphi_{y})}{4\tau}}\cosh{\left(\frac{1}{\tau}\sqrt{\varphi_{x}^{2}-\varphi_{x}\varphi_{y}+\varphi_{y}^{2}}\right)}(-\varphi_{x}-\varphi_{y})^{4}}{8\tau^{4}(\varphi_{x}^{2}-\varphi_{x}\varphi_{y}+\varphi_{y})^{2}} \right.\right.
+\left.\left.\frac{3e^{\frac{5(\varphi_{x}+\varphi_{y})}{4\tau}}\sinh{\left(\frac{1}{\tau}\sqrt{\varphi_{x}^{2}-\varphi_{x}\varphi_{y}+\varphi_{y}^{2}}\right)}(-\varphi_{x}-\varphi_{y})^{3}}{2\tau^{2}(\varphi_{x}^{2}-\varphi_{x}\varphi_{y}+\varphi_{y})^{\frac{3}{2}}} \right.\right. \nonumber \\
&+&\left.\left.\frac{15e^{\frac{5(\varphi_{x}+\varphi_{y})}{4\tau}}\cosh{\left(\frac{1}{\tau}\sqrt{\varphi_{x}^{2}-\varphi_{x}\varphi_{y}+\varphi_{y}^{2}}\right)}(-\varphi_{x}-\varphi_{y})^{4}}{8\tau^{2}(\varphi_{x}^{2}-\varphi_{x}\varphi_{y}+\varphi_{y})^{3}} \right.\right.
-\left.\left.\frac{15e^{\frac{5(\varphi_{x}+\varphi_{y})}{4\tau}}\sinh{\left(\frac{1}{\tau}\sqrt{\varphi_{x}^{2}-\varphi_{x}\varphi_{y}+\varphi_{y}^{2}}\right)}(-\varphi_{x}-\varphi_{y})^{4}}{8\tau(\varphi_{x}^{2}-\varphi_{x}\varphi_{y}+\varphi_{y})^{\frac{7}{2}}} \right.\right. \nonumber \\
&-&\left.\left.\frac{3e^{\frac{5(\varphi_{x}+\varphi_{y})}{4\tau}}\sinh{\left(\frac{1}{\tau}\sqrt{\varphi_{x}^{2}-\varphi_{x}\varphi_{y}+\varphi_{y}^{2}}\right)}(\varphi_{x}+\varphi_{y})^{3}}{2\tau^{2}(\varphi_{x}^{2}-\varphi_{x}\varphi_{y}+\varphi_{y})^{\frac{5}{2}}} \right.\right.
+\left.\left.\frac{15e^{\frac{5(\varphi_{x}+\varphi_{y})}{4\tau}}\cosh{\left(\frac{1}{\tau}\sqrt{\varphi_{x}^{2}-\varphi_{x}\varphi_{y}+\varphi_{y}^{2}}\right)}(\varphi_{x}+\varphi_{y})^{4}}{8\tau^{2}(\varphi_{x}^{2}-\varphi_{x}\varphi_{y}+\varphi_{y})^{3}} \right.\right.\nonumber \\
&-&\left.\left.\frac{15e^{\frac{5(\varphi_{x}+\varphi_{y})}{4\tau}}\sinh{\left(\frac{1}{\tau}\sqrt{\varphi_{x}^{2}-\varphi_{x}\varphi_{y}+\varphi_{y}^{2}}\right)}(\varphi_{x}+\varphi_{y})^{4}}{8\tau(\varphi_{x}^{2}-\varphi_{x}\varphi_{y}+\varphi_{y})^{\frac{7}{2}}} \right.\right.
+\left.\left.\frac{9e^{\frac{5(\varphi_{x}+\varphi_{y})}{4\tau}}\sinh{\left(\frac{1}{\tau}\sqrt{\varphi_{x}^{2}-\varphi_{x}\varphi_{y}+\varphi_{y}^{2}}\right)}(-\varphi_{x}-\varphi_{y})^{2}}{4\tau^{3}(\varphi_{x}^{2}-\varphi_{x}\varphi_{y}+\varphi_{y})^{\frac{3}{2}}} \right.\right. \nonumber \\
&-&\left.\left.\frac{3e^{\frac{5(\varphi_{x}+\varphi_{y})}{4\tau}}\cosh{\left(\frac{1}{\tau}\sqrt{\varphi_{x}^{2}-\varphi_{x}\varphi_{y}+\varphi_{y}^{2}}\right)}(-\varphi_{x}-\varphi_{y})^{3}}{2\tau^{3}(\varphi_{x}^{2}-\varphi_{x}\varphi_{y}+\varphi_{y})^{2}} \right.\right.
-\left.\left.\frac{6e^{\frac{5(\varphi_{x}+\varphi_{y})}{4\tau}}\sinh{\left(\frac{1}{\tau}\sqrt{\varphi_{x}^{2}-\varphi_{x}\varphi_{y}+\varphi_{y}^{2}}\right)}(-\varphi_{x}-\varphi_{y})}{\tau^{2}(\varphi_{x}^{2}-\varphi_{x}\varphi_{y}+\varphi_{y})^{\frac{3}{2}}} \right.\right.\nonumber \\
&-&\left.\left.\frac{3e^{\frac{5(\varphi_{x}+\varphi_{y})}{4\tau}}\sinh{\left(\frac{1}{\tau}\sqrt{\varphi_{x}^{2}-\varphi_{x}\varphi_{y}+\varphi_{y}^{2}}\right)}(-\varphi_{x}-\varphi_{y})^{3}}{4\tau^{3}(\varphi_{x}^{2}-\varphi_{x}\varphi_{y}+\varphi_{y})^{\frac{5}{2}}} \right.\right.
-\left.\left.\frac{9e^{\frac{5(\varphi_{x}+\varphi_{y})}{4\tau}}\cosh{\left(\frac{1}{\tau}\sqrt{\varphi_{x}^{2}-\varphi_{x}\varphi_{y}+\varphi_{y}^{2}}\right)}(-\varphi_{x}-\varphi_{y})^{2}}{\tau^{2}(\varphi_{x}^{2}-\varphi_{x}\varphi_{y}+\varphi_{y})^{2}} \right.\right.\nonumber \\
&+&\left.\left.\frac{9e^{\frac{5(\varphi_{x}+\varphi_{y})}{4\tau}}\sinh{\left(\frac{1}{\tau}\sqrt{\varphi_{x}^{2}-\varphi_{x}\varphi_{y}+\varphi_{y}^{2}}\right)}(-\varphi_{x}-\varphi_{y})^{2}}{\tau(\varphi_{x}^{2}-\varphi_{x}\varphi_{y}+\varphi_{y})^{\frac{5}{2}}} \right.\right.
+\left.\left.\frac{9e^{\frac{5(\varphi_{x}+\varphi_{y})}{4\tau}}\sinh{\left(\frac{1}{\tau}\sqrt{\varphi_{x}^{2}-\varphi_{x}\varphi_{y}+\varphi_{y}^{2}}\right)}(\varphi_{x}+\varphi_{y})^{2}}{\tau^{3}(\varphi_{x}^{2}-\varphi_{x}\varphi_{y}+\varphi_{y})^{\frac{3}{2}}} \right.\right.\nonumber \\
&+&\left.\left.\frac{3e^{\frac{5(\varphi_{x}+\varphi_{y})}{4\tau}}\cosh{\left(\frac{1}{\tau}\sqrt{\varphi_{x}^{2}-\varphi_{x}\varphi_{y}+\varphi_{y}^{2}}\right)}(\varphi_{x}+\varphi_{y})^{3}}{\tau^{3}(\varphi_{x}^{2}-\varphi_{x}\varphi_{y}+\varphi_{y})^{2}} \right.\right.
+\left.\left.\frac{6e^{\frac{5(\varphi_{x}+\varphi_{y})}{4\tau}}\sinh{\left(\frac{1}{\tau}\sqrt{\varphi_{x}^{2}-\varphi_{x}\varphi_{y}+\varphi_{y}^{2}}\right)}(\varphi_{x}+\varphi_{y})}{\tau^{2}(\varphi_{x}^{2}-\varphi_{x}\varphi_{y}+\varphi_{y})^{\frac{3}{2}}} \right.\right.\nonumber \\
&-&\left.\left.\frac{3e^{\frac{5(\varphi_{x}+\varphi_{y})}{4\tau}}\sinh{\left(\frac{1}{\tau}\sqrt{\varphi_{x}^{2}-\varphi_{x}\varphi_{y}+\varphi_{y}^{2}}\right)}(\varphi_{x}+\varphi_{y})^{4}}{4\tau^{3}(\varphi_{x}^{2}-\varphi_{x}\varphi_{y}+\varphi_{y})^{\frac{5}{2}}} \right.\right.
-\left.\left.\frac{9e^{\frac{5(\varphi_{x}+\varphi_{y})}{4\tau}}\cosh{\left(\frac{1}{\tau}\sqrt{\varphi_{x}^{2}-\varphi_{x}\varphi_{y}+\varphi_{y}^{2}}\right)}(\varphi_{x}+\varphi_{y})^{2}}{\tau^{2}(\varphi_{x}^{2}-\varphi_{x}\varphi_{y}+\varphi_{y})^{2}} \right.\right.\nonumber \\
&+&\left.\left.\frac{9e^{\frac{5(\varphi_{x}+\varphi_{y})}{4\tau}}\sinh{\left(\frac{1}{\tau}\sqrt{\varphi_{x}^{2}-\varphi_{x}\varphi_{y}+\varphi_{y}^{2}}\right)}(\varphi_{x}+\varphi_{y})^{2}}{\tau(\varphi_{x}^{2}-\varphi_{x}\varphi_{y}+\varphi_{y})^{\frac{5}{2}}}\right]\right] .
\end{eqnarray}

\end{document}